\documentclass[11pt]{article}

\usepackage{cite}
\usepackage{hyperref}
\usepackage{cancel}

\usepackage{lineno}

\usepackage{amsmath}
\usepackage{amssymb,amsfonts,amsthm}
\usepackage{epsfig,graphicx}
\usepackage{url} 
\usepackage{bm} 
\usepackage{slashed}
\usepackage{cite}

\usepackage{color}
\usepackage{pstricks}

\usepackage{rotating}

\usepackage{fancyhdr}
\usepackage{textcomp}
\usepackage{subcaption}

\usepackage{color}
\usepackage{axodraw4j}
\usepackage{pstricks}
\usepackage{graphicx}
\usepackage{float} 
\usepackage[utf8]{inputenc}

\newcommand{\unit}{\leavevmode\hbox{\small1\kern-3.6pt\normalsize1}}

\newcommand{\be}{\begin{equation}}
\newcommand{\ee}{\end{equation}}

\newcommand{\bea}{\begin{eqnarray}}
\newcommand{\eea}{\end{eqnarray}}

\def\simlt{\stackrel{<}{{}_\sim}}
\def\simgt{\stackrel{>}{{}_\sim}}

\def\sigmavh{\langle\sigma v\rangle_0}
\def\sigsi{\sigma^{SI}_{S_1 p}}

\def\moi{m_{i}}
\def\msl{m_{S_1}}
\def\msh{m_{S_2}}
\def\mi{m_{S_i}}

\begin{document}

\title{Reopening the Higgs portal for Single Scalar Dark Matter}

\thispagestyle{empty}
\begin{flushright}

 {\small 
 IFT-UAM/CSIC-16-113; 
 IPPP/16/107; 
 DCTP/16/214}\\

  \vspace*{2.mm}{\today}
\end{flushright}

\begin{center}
  {\bf {\LARGE Reopening the Higgs Portal for Singlet Scalar Dark Matter}}

\renewcommand*{\thefootnote}{\fnsymbol{footnote}}
\setcounter{footnote}{3}

  \vspace{0.5cm}
  {\large
  J.~A.~Casas $^{a}$,
    D.~G.~Cerde\~no $^{b,a}$,
    J.~M.~Moreno $^{a}$ 
   and
    J. Quilis $^{a}$
  }
  \\[0.2cm] 

  {\footnotesize{
       $^a$ Instituto de F\'{\i}sica Te\'{o}rica UAM/CSIC, Universidad Aut\'{o}noma de Madrid,  28049, Madrid, Spain\\
       $^b$ Institute for Particle Physics Phenomenology, Department of Physics\\
	Durham University, Durham DH1 3LE, United Kingdom\\
        }
    }

\vspace*{0.7cm}

  \begin{abstract}
A real singlet scalar, connected to the Standard Model sector through a portal with the Higgs boson, is one of the simplest and most popular models for dark matter (DM). However, the experimental advances in direct and indirect DM searches, together with the latest results from the LHC, have ruled out vast areas of the parameter space of this scenario; and are expected to probe it completely within the next years, ruling it out if no signal is found.
Motivated by the simplicity of this model, in this article we address a minimal, renormalizable extension 
that could evade detection, consisting of the addition of an extra real singlet scalar field in the dark sector.
We analyze the physical constraints on the model and show that the new annihilation and/or coannihilation channels involving the extra singlet allow to reproduce the correct DM relic abundance while avoiding the bounds from direct and indirect searches 
for any DM mass above 50~GeV.
We also show that, in some interesting regions of the parameter space, the extra particle can be integrated-out, leaving a ``clever" effective theory (just involving the DM particle and the Higgs), that essentially reproduces the results.
   \end{abstract}
\end{center}

\newpage
\pagestyle{plain}


\section{Introduction}
\label{sec:introduction}

The nature of more than 80\% of the matter in our Universe is still unknown. Over the past century, substantial evidence has been collected from astrophysical and cosmological observations that supports the existence of a new type of {\em dark} matter (DM), that does not emit or absorb light, and that cannot be explained by the Standard Model (SM). This window to new physics is currently being thoroughly probed by dedicated direct and indirect DM experiments, as well as by the Large Hadron Collider (LHC), with increasing sensitivities.

Among the many particle physics candidates for DM, the singlet-scalar Higgs portal (SHP) model stands out as one of the most economical and popular scenarios. \cite{Silveira:1985rk,McDonald:1993ex,Burgess:2000yq}.
It simply consists of one extra singlet scalar, $S$ (the DM particle), which is minimally coupled to the SM through interactions with the ordinary Higgs (the only ones allowed at the renormalizable level). The corresponding Lagrangian reads 
\begin{equation}
\mathcal{L}_{\rm SHP}=\mathcal{L}_{\rm SM}+\frac{1}{2}\partial_{\mu} S \partial^{\mu} S- \frac{1}{2}m_0^2 S^2-\frac{1}{2}\lambda_S |H|^2 S^2 -\frac{1}{4!}\lambda_{4} S^4 .
\label{HPlagr}
\end{equation}
In the previous equation $S$ has been assumed to be a real field, but the modification for  the complex case is trivial.
Furthermore, a discrete symmetry $S\rightarrow -S$ has been imposed in order to ensure the stability of the DM particle; apart from this, the above renormalizable Lagrangian is completely general. After electroweak (EW) symmetry breaking, the Higgs field acquires a vacuum expectation value, $H^0=(v+h)/\sqrt{2}$, and new terms appear, including a trilinear coupling between $S$ and the Higgs boson, $(\lambda_S v/2) h S^2$. 
The phenomenology of this model has been explored in other contexts as well \cite{Davoudiasl:2004be, Barger:2007im, Lerner:2009xg, Grzadkowski:2009mj}.

\begin{figure}[t!]
\centering 
\begin{center}
\fcolorbox{white}{white}{
  \begin{picture}(464,99) (36,-9)
    \SetWidth{1.0}
    \SetColor{Black}
    \Line[dash,dashsize=5,arrow,arrowpos=0.5,arrowlength=5,arrowwidth=2,arrowinset=0.2](48,67)(80,35)
    \Line[dash,dashsize=5,arrow,arrowpos=0.5,arrowlength=5,arrowwidth=2,arrowinset=0.2](48,3)(80,35)
    \Line[dash,dashsize=5,arrow,arrowpos=0.5,arrowlength=5,arrowwidth=2,arrowinset=0.2](80,35)(128,35)
    \Line[arrow,arrowpos=0.5,arrowlength=5,arrowwidth=2,arrowinset=0.2](128,35)(160,67)
    \Line[arrow,arrowpos=0.5,arrowlength=5,arrowwidth=2,arrowinset=0.2](128,35)(160,3)
    \Line[dash,dashsize=5,arrow,arrowpos=0.5,arrowlength=5,arrowwidth=2,arrowinset=0.2](240,67)(272,35)
    \Line[dash,dashsize=5,arrow,arrowpos=0.5,arrowlength=5,arrowwidth=2,arrowinset=0.2](240,3)(272,35)
    \Line[dash,dashsize=5,arrow,arrowpos=0.5,arrowlength=5,arrowwidth=2,arrowinset=0.2](272,35)(304,67)
    \Line[dash,dashsize=5,arrow,arrowpos=0.5,arrowlength=5,arrowwidth=2,arrowinset=0.2](272,35)(304,3)
    \Line[dash,dashsize=5,arrow,arrowpos=0.5,arrowlength=5,arrowwidth=2,arrowinset=0.2](368,67)(416,67)
    \Line[dash,dashsize=5,arrow,arrowpos=0.5,arrowlength=5,arrowwidth=2,arrowinset=0.2](416,67)(464,67)
    \Line[dash,dashsize=5,arrow,arrowpos=0.5,arrowlength=5,arrowwidth=2,arrowinset=0.2](416,67)(416,3)
    \Line[dash,dashsize=5,arrow,arrowpos=0.5,arrowlength=5,arrowwidth=2,arrowinset=0.2](368,3)(416,3)
    \Line[dash,dashsize=5,arrow,arrowpos=0.5,arrowlength=5,arrowwidth=2,arrowinset=0.2](416,3)(464,3)
    \Text(33,67)[lb]{\Large{\Black{$S$}}}
    \Text(33,-13)[lb]{\Large{\Black{$S$}}}
    \Text(90,13)[lb]{\Large{\Black{$h$}}}
    \Text(160,69)[lb]{\Large{\Black{$SM$}}}
    \Text(161,-12)[lb]{\Large{\Black{$SM$}}}
    \Text(224,67)[lb]{\Large{\Black{$S$}}}
    \Text(225,-12)[lb]{\Large{\Black{$S$}}}
    \Text(304,67)[lb]{\Large{\Black{$h$}}}
    \Text(304,-12)[lb]{\Large{\Black{$h$}}}
    \Text(351,-14)[lb]{\Large{\Black{$S$}}}
    \Text(353,66)[lb]{\Large{\Black{$S$}}}
    \Text(390,29)[lb]{\Large{\Black{$S$}}}
    \Text(465,66)[lb]{\Large{\Black{$h$}}}
    \Text(463,-14)[lb]{\Large{\Black{$h$}}}
  \end{picture}
}
\end{center}
\caption{Singlet-scalar Higgs portal scenario (SHP): annihilation processes of the DM candidate, $S$. 
}
\label{fig:HPprocesses}
\end{figure}
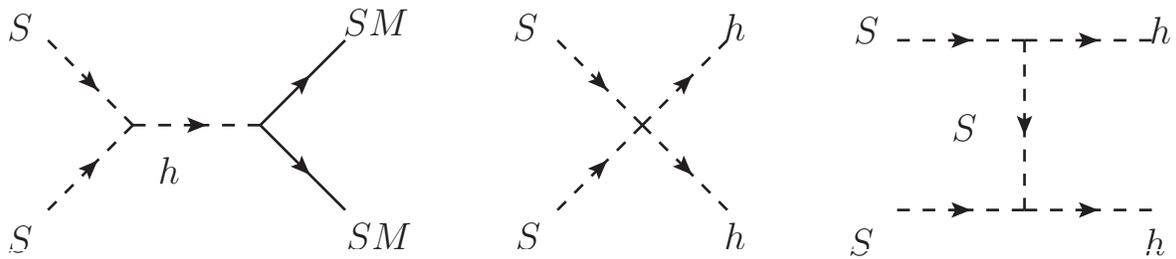

Assuming that the $S-$particles are in thermal equilibrium in the early universe, the final DM relic density is determined by their primordial annihilation rate into SM-particles. The relevant processes, illustrated in Fig.\,\ref{fig:HPprocesses}, are usually dominated by the $s-$channel annihilation through a Higgs boson  (leftmost diagram of the figure).

\begin{figure}[t!]
\centering 
\includegraphics[width=0.49\linewidth]{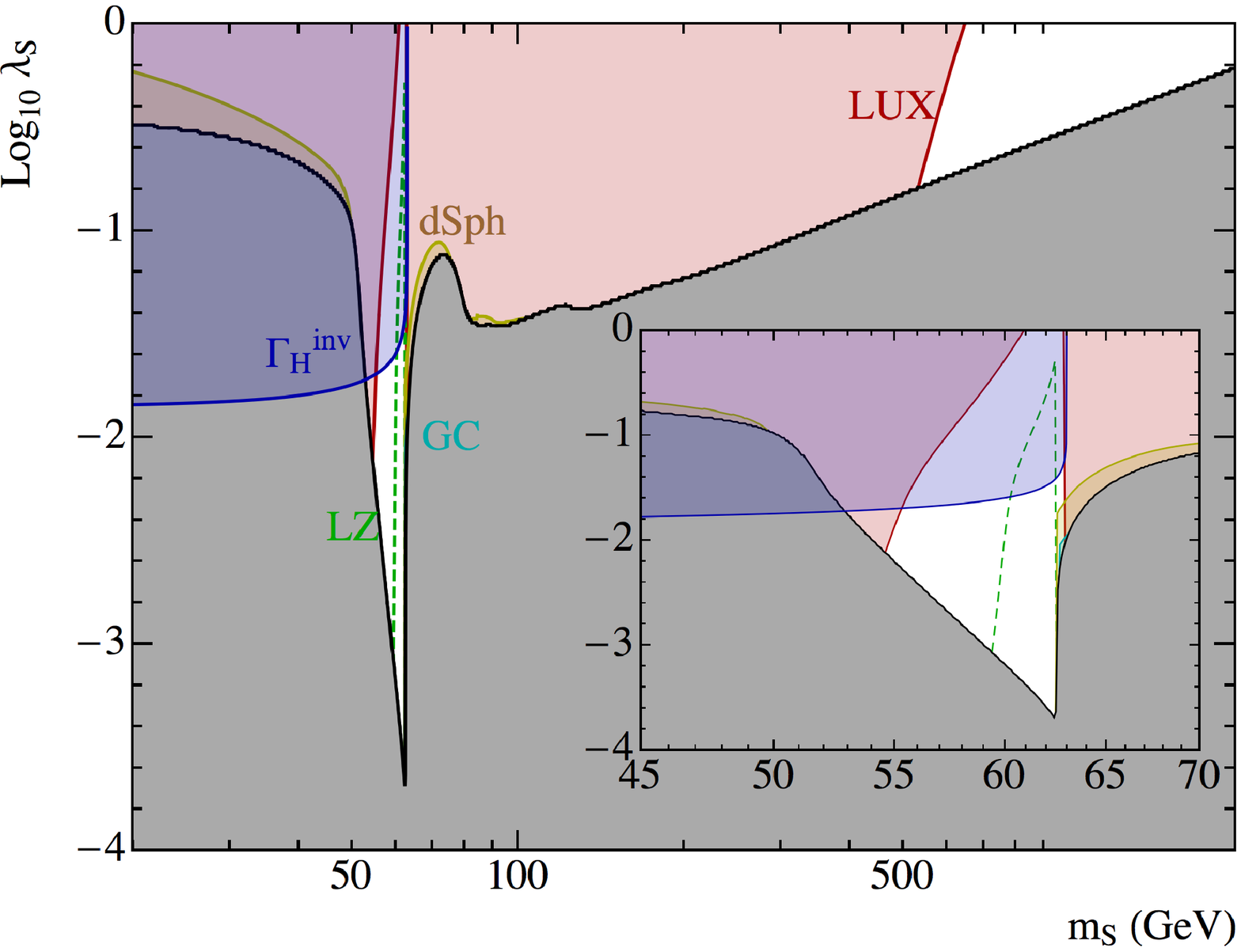}
\includegraphics[width=0.49\linewidth]{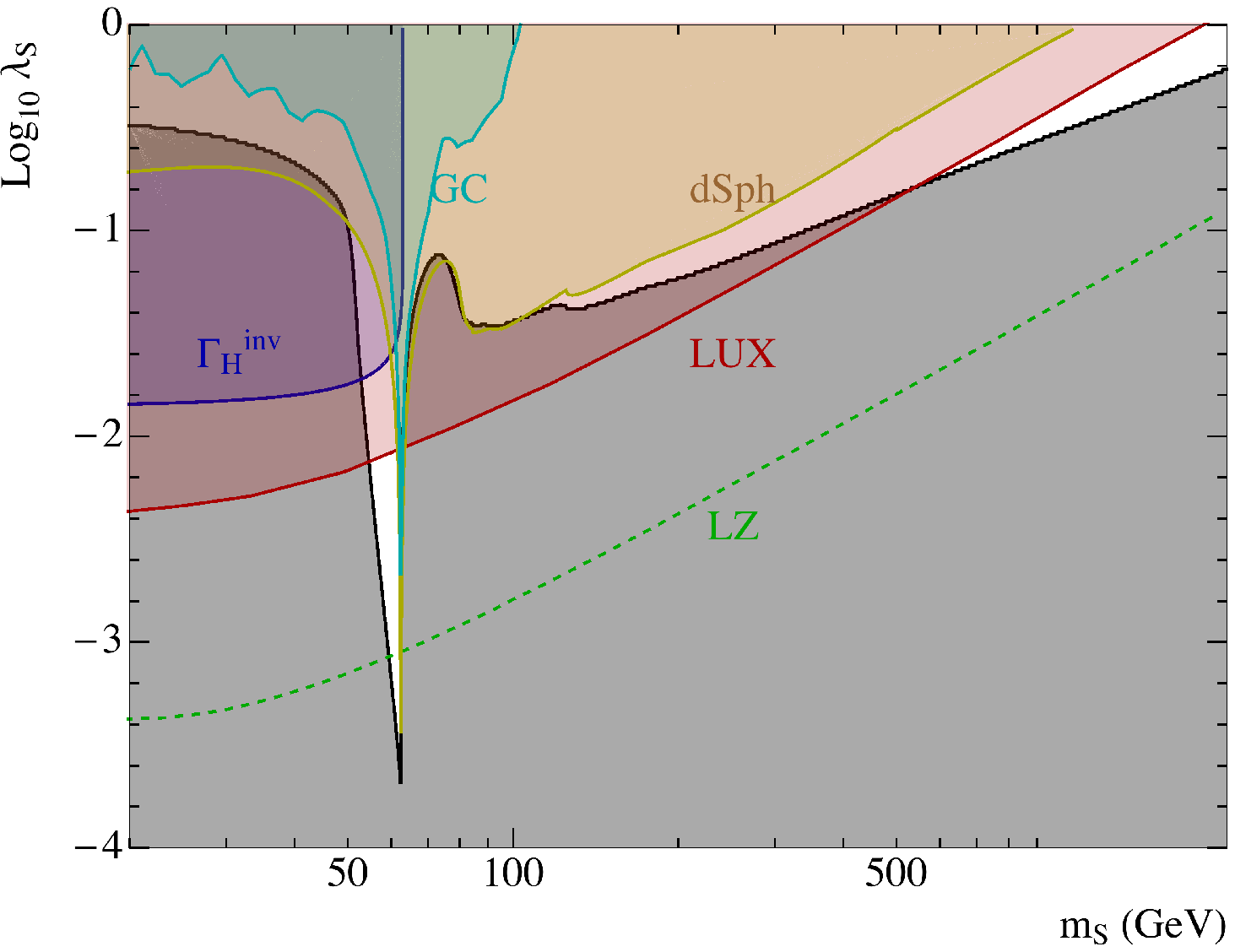}

\caption{
Excluded regions on the parameter space of the SHP model from different experimental constraints. The gray area (below the black line) is excluded since the relic density exceeds the Planck result. The blue area (labeled $\Gamma_H^{inv}$) is ruled out from the invisible Higgs width. The red area (LUX) is excluded by direct DM detection limits. Yellow (dSph) and cyan (GC) areas are excluded by indirect detection constraints on the continuum spectrum of gamma-rays (from dwarf Spheroidal galaxies) and monochromatic gamma-ray lines (from the Galactic Centre), respectively. The dashed green line represents the predicted reach of the future LZ detector. The left panel includes a scale factor, $\xi$, in the calculations while in the right plot it is assumed that some extra non-thermal effects amend the prediction for the relic density, so that  $\xi=1$. 
}
\label{fig:HPc}
\end{figure}

The efficiency of the annihilation depends on just two parameters, $\{m_0, \lambda_S\}$ or, equivalently, $\{m_S, \lambda_S\}$, where $m_S^2=m_0^2+\lambda_Sv^2/2$ is the physical $S-$mass after EW breaking. Fig.\,\ref{fig:HPc} 
shows the (black) line in the $\{m_S, \lambda_S\}$ plane along which
the relic abundance of $S$, $\Omega_Sh^2$, coincides with the Planck result $\Omega_{CDM}h^2=0.1198\pm 0.003$ at $2\sigma$ \cite{Ade:2015xua}. The (gray) region below is in principle excluded, as it corresponds to a higher relic density.

The model is subject to a number of experimental and observational constraints, which rule out large regions of the parameter space. These include limits from direct detection experiments  \cite{Feldstein:2009tr, Chang:2008xa, Dienes:2013xya, Fan:2010gt, Dent:2016iht, Gluscevic:2015sqa, Kumar:2013iva, Savage:2008er, Billard:2013qya, He:2008qm, Farina:2009ez, Bandyopadhyay:2010cc, He:2016mls}, indirect searches \cite{Bonnivard:2014kza, Ichikawa:2016nbi, Klop:2016lug, Ipek:2014gua, Daylan:2014rsa, Hooper:2010mq, Hooper:2011ti, Goodenough:2009gk, Abazajian:2012pn, Gordon:2013vta, TheFermi-LAT:2015kwa, Berlin:2016gtr, deSimone:2014pda, Hooper:2012sr, Ackermann:2015zua, Giesen:2015ufa, Cirelli:2013hv, Berlin:2014tja, Duerr:2015bea, Beniwal:2015sdl, Profumo:2010kp,Sage:2016xkb, Duerr:2015mva, Cuoco:2016jqt, Balazs:2014jla,Duerr:2015aka,Han:2015dua}, as well as collider bounds \cite{Kozaczuk:2015bea, He:2007tt, Djouadi:2012zc, Craig:2014lda, Khachatryan:2016whc, Ko:2016xwd, Han:2016gyy, Carpenter:2013xra}.  
We illustrate the effects of these limits in Fig.\,\ref{fig:HPc}.
In deriving direct and indirect detection bounds, we are assuming by default (left panel) that the density of $S$ scales up in the same way as its cosmological relic abundance. Thus, we consider a scale factor $\xi\equiv\Omega_{S}/\Omega_{CDM}$ for direct detection and $\xi^2$ for indirect detection. In the region where $\xi<1$, $S$ cannot be the only DM component, so contributions from other particles (e.g., axions) are needed. 
The region where $\xi>1$ (gray area) is obviously excluded (though perhaps could be rescued if some non-standard cosmology is invoked, see below). For this reason, we have not showed the shadowed regions inside this gray area. It is worth noting that the excluded areas are extremely sensitive to astrophysical uncertainties in the DM halo parameters \cite{Benito:2016kyp} and nuclear uncertainties in the hadronic matrix elements \cite{Duerr:2015aka}.

Current bounds from direct DM detection, most notably from the new results from LUX \cite{Akerib:2016vxi} and PandaX-II \cite{Tan:2016zwf}, 
set an upper bound on the DM-nucleon elastic scattering cross section (and hence on the DM coupling to the Higgs). This rules out the red area in Fig.\,\ref{fig:HPc}. Next-generation experiments, with larger targets and improved sensitivity are going to further explore this parameter space. We indicate in the figure the expected reach of the LZ detector by means of a green dashed line.
Similarly, Fermi-LAT data on the continuum gamma-ray flux from dwarf spheroidal galaxies (dSPh) and monochromatic gamma-ray lines from the Galactic Centre set upper bounds on the DM annihilation cross section which also rule out some areas of the parameter space, mainly for DM masses below 100~GeV (light brown and cyan areas respectively). 
It should be noticed that, as $\lambda_S$ decreases, the $\xi-$factor increases, so that the indirect detection rate increases as well. Consequently, the excluded areas from indirect detection extend downwards in the plot.
Finally, for masses below $\sim63$~GeV, the DM can contribute to the invisible decay of the SM Higgs boson. Current LHC constraints on this quantity set an upper bound on the DM-Higgs coupling \cite{Khachatryan:2016whc}. The blue region in Fig.\,\ref{fig:HPc} is excluded for this reason.

For comparison, the right panel of Fig.\,\ref{fig:HPc} shows the direct and indirect detection constraints when the local DM density is assumed to take the canonical value, $\rho_0=0.3$~GeV~cm$^{-3}$, regardless of the computed thermal relic abundance; in other words, we have set $\xi=1$. 
This would apply if non-thermal effects modified the final relic abundance, reconciling it with the observed one (see, e.g., Ref.\,\cite{Gelmini:2006pw}). 
Note that, since the value of $\xi$ has been fixed, the areas excluded by indirect detection bounds now extend upwards.

In either case, the conclusion is that the combination of experimental constraints
and the requirement of obtaining the correct relic abundance rules out a big and interesting portion of the viable parameter space of the Higgs portal (see Ref. \cite{Escudero:2016gzx} for a recent comprehensive study), leaving only the white areas in Fig.\,\ref{fig:HPc}. 
Interestingly, as previous analyses have shown \cite{Cline:2013gha,Queiroz:2014yna, Feng:2014vea,Wu:2016mbe}
there still remains a narrow window of $S-$masses in the Higgs-funnel region
($m_S \simeq m_h/2$) and, besides, there is a large allowed range for higher masses, $m_S\simgt 500$ GeV. Next generation experiments such as XENON1T \cite{Aprile:2015uzo} and, especially, LZ \cite{Akerib:2015cja} (shown explicitly) will test completely the region of large DM masses and a large part of the narrow window at the Higgs-resonance. In particular, LZ could exclude the Higgs-portal scenario almost completely, or, hopefully, get a positive detection. The possibility of totally closing the Higgs-portal windows in the near future using complementary constraints from indirect detection has been analyzed in refs.~\cite{Cline:2013gha,Feng:2014vea,Escudero:2016gzx}.

Various solutions have been proposed in order to avoid experimental constraints in the SHP model. In general, in order to break the correlation between the relic abundance and direct detection predictions, the model has to be extended. For example, the mediator (Higgs) sector can be enlarged 
with new scalars \cite{Ishiwata:2011aa,Abada:2011qb, Modak:2013jya}.
Non-linear Higgs portals \cite{Brivio:2015kia} and  
high-dimensional operators in models with composite Higgs \cite{Fonseca:2015gva} have been considered as well.
One can also extend the dark sector to include new particles charged under the SM gauge group, such as a doublet, a triplet, or a top-partner (see, e.g.,  \cite{Cohen:2011ec, Cheung:2013dua, Giacchino:2015hvk, Baek:2016lnv}), or even consider multicomponent dark matter scenarios \cite{Belanger:2011ww,Biswas:2013nn,Modak:2013jya,Bhattacharya:2016ysw,Drozd:2011aa}.
More complex scenarios have also been analysed, where both the dark matter and mediator sectors are enlarged \cite{Hierro:2016nwm}, for example, adding new portals related to neutrino physics \cite{Kawana:2014zxa,Escudero:2016tzx,Escudero:2016ksa,Bhattacharya:2016qsg}.
There is also the possibility that the dark matter is a singlet-fermion, in which case the Higgs-portal interactions occur at  the non-renormalizable level.
Finally, one can consider changing the nature of the DM candidate, see for example Refs.\,\cite{Fischer:2011zz,Arcadi:2016qoz}.

The goal of this paper is to consider and examine the most economical modification of the conventional SHP model that could escape the present and future searches, thus offering a viable (slightly modified) Higgs-portal scenario if a positive detection does not occur.
The model consists of the addition of a second singlet scalar in the dark sector, which opens up new annihilation and coannihilation channels (previous work in this line has been carried out in Ref.~\cite{Ghorbani:2014gka}).

The article is organised as follows.
The model is introduced in Section\,\ref{sec:extended-portal}, 
where we explain how the correct relic abundance can be achieved for large regions of the parameter space. 
In Section\,\ref{sec:constraints}, we describe the 
various experimental constraints to which the model is subject, and explain the way we have evaluated them. They include bounds from direct and indirect DM detection, the lifetime of the extra particle and the invisible decay of the Higgs boson.
In Section\,\ref{sec:results} we perform a scan in the parameter space, explicitly showing that our model is viable for any DM mass above 50~GeV, thereby reopening the Higgs portal for scalar DM.
In Section\,\ref{sec:EFT} we discuss the interpretation of this model in terms of an Effective Field Theory.
Finally, the conclusions of our study are presented in Section\,\ref{sec:conclusions}.
The Appendix is devoted to the calculation of the relevant radiative corrections for DM processes.

\section{The extended singlet-scalar Higgs portal (ESHP)}
\label{sec:extended-portal}

The modification of the conventional SHP model that we consider consists simply of extending the DM sector with the addition of a second scalar. Denoting $S_1$, $S_2$ the two scalar particles, and imposing a global $Z_2$ symmetry ($S_1\rightarrow -S_1$, $S_2\rightarrow -S_2$) in order to guarantee the stability of the lightest one, the most general renormalizable Lagrangian reads
\begin{eqnarray}
\mathcal{L}_{\rm ESHP}&=&\mathcal{L}_{\rm SM}+\frac{1}{2}\sum_{i=1,2}\left[(\partial_{\mu} S_i)^2 -\moi^2 S_i^2-\frac{1}{12}\lambda_{i4}S_i^4\right]
-\frac{1}{6}\lambda_{13} S_1 S_2^3 -\frac{1}{6}\lambda_{31} S_1^3 S_2-\frac{1}{4}\lambda_{22} S_1^2 S_2^2
\nonumber\\
&&-\frac{1}{2}\lambda_1S_1^2|H|^2-\frac{1}{2}\lambda_2S_2^2|H|^2-\lambda_{12} S_1 S_2 \left(|H|^2 - \frac{v^2}{2}\right)\ ,
\label{HPextlagr}
\end{eqnarray}
where the subscript ESHP stands for ``extended singlet-scalar Higgs portal".
The terms in the second line describe the DM/SM interactions, which occur through the Higgs sector. After EW breaking, $H^0=(v+h)/\sqrt{2}$, there appear new terms, including trilinear terms between $S_{1,2}$ and the Higgs boson, such as $(\lambda_{12} v)\, h S_1S_2$. Stability constraints in this type of models have been studied in Ref.\,\cite{Kannike:2016fmd}. We have chosen $S_1, S_2$ to be 
 the final mass eigenstates (after EW breaking), with physical masses, $\mi^2 = \moi^2 + \lambda_i v^2/2$, thus the form of the last term in eq.(\ref{HPextlagr}). 
From now on, $S_1$ will represent the lightest mass eigenstate of the dark sector, and thus the DM particle.

\subsection{The relic density}
\label{sec:relicdensity}

The extra terms in the Lagrangian open up new ways of DM annihilation, illustrated in Fig.~\ref{fig:EHPprocesses}. These include processes mediated by $S_2$ (in $t-$channel) and co-annihilation processes. Besides, if $S_1$ and $S_2$ are in thermal equilibrium between them (thanks to the interaction terms in the first line of eq.(\ref{HPextlagr})), the processes driving $S_2-$annihilation contribute to the DM annihilation as well.

We are interested in the possibility that $S_1$ plays the role of DM, and that it reproduces the observed relic density while evading the bounds discussed in the previous section for the usual SHP model. Hence we will mainly focus in the regime where $\lambda_1$ (the equivalent to $\lambda_S$ in the ordinary Higgs-portal) is small. As a matter of fact, $\lambda_1$ might be even vanishing, and the processes of Fig.\,\ref{fig:EHPprocesses} could still produce the necessary annihilation. However, this is not a natural choice from the point of view of quantum field theory. Since 1-loop diagrams with two $\lambda_{12}$ vertices generate $S_1^2|H|^2$ interactions, a conservative attitude is to assume that $\lambda_1$ is not smaller than $\sim \lambda_{12}^2/(4\pi)^2$. 
The same argument holds for $\lambda_2$. Actually, for the sake of definiteness we will set $\lambda_2= \lambda_{12}^2/(4\pi)^2$ through the paper.

\begin{figure}[t!]
\begin{center}
\fcolorbox{white}{white}{
  \begin{picture}(464,99) (36,-9)
    \SetWidth{1.0}
    \SetColor{Black}
    \Line[dash,dashsize=5,arrow,arrowpos=0.5,arrowlength=5,arrowwidth=2,arrowinset=0.2](48,67)(80,35)
    \Line[dash,dashsize=5,arrow,arrowpos=0.5,arrowlength=5,arrowwidth=2,arrowinset=0.2](48,3)(80,35)
    \Line[dash,dashsize=5,arrow,arrowpos=0.5,arrowlength=5,arrowwidth=2,arrowinset=0.2](80,35)(128,35)
    \Line[arrow,arrowpos=0.5,arrowlength=5,arrowwidth=2,arrowinset=0.2](128,35)(160,67)
    \Line[arrow,arrowpos=0.5,arrowlength=5,arrowwidth=2,arrowinset=0.2](128,35)(160,3)
    \Line[dash,dashsize=5,arrow,arrowpos=0.5,arrowlength=5,arrowwidth=2,arrowinset=0.2](240,67)(272,35)
    \Line[dash,dashsize=5,arrow,arrowpos=0.5,arrowlength=5,arrowwidth=2,arrowinset=0.2](240,3)(272,35)
    \Line[dash,dashsize=5,arrow,arrowpos=0.5,arrowlength=5,arrowwidth=2,arrowinset=0.2](272,35)(304,67)
    \Line[dash,dashsize=5,arrow,arrowpos=0.5,arrowlength=5,arrowwidth=2,arrowinset=0.2](272,35)(304,3)
    \Line[dash,dashsize=5,arrow,arrowpos=0.5,arrowlength=5,arrowwidth=2,arrowinset=0.2](368,67)(416,67)
    \Line[dash,dashsize=5,arrow,arrowpos=0.5,arrowlength=5,arrowwidth=2,arrowinset=0.2](416,67)(464,67)
    \Line[dash,dashsize=5,arrow,arrowpos=0.5,arrowlength=5,arrowwidth=2,arrowinset=0.2](416,67)(416,3)
    \Line[dash,dashsize=5,arrow,arrowpos=0.5,arrowlength=5,arrowwidth=2,arrowinset=0.2](368,3)(416,3)
    \Line[dash,dashsize=5,arrow,arrowpos=0.5,arrowlength=5,arrowwidth=2,arrowinset=0.2](416,3)(464,3)
    \Text(33,67)[lb]{\Large{\Black{$S_i$}}}
    \Text(33,-13)[lb]{\Large{\Black{$S_j$}}}
    \Text(90,13)[lb]{\Large{\Black{$h$}}}
    \Text(160,69)[lb]{\Large{\Black{$SM$}}}
    \Text(161,-12)[lb]{\Large{\Black{$SM$}}}
    \Text(224,67)[lb]{\Large{\Black{$S_i$}}}
    \Text(225,-12)[lb]{\Large{\Black{$S_j$}}}
    \Text(304,67)[lb]{\Large{\Black{$h$}}}
    \Text(304,-12)[lb]{\Large{\Black{$h$}}}
    \Text(351,-14)[lb]{\Large{\Black{$S_i$}}}
    \Text(353,66)[lb]{\Large{\Black{$S_j$}}}
    \Text(390,29)[lb]{\Large{\Black{$S_k$}}}
    \Text(465,66)[lb]{\Large{\Black{$h$}}}
    \Text(463,-14)[lb]{\Large{\Black{$h$}}}
  \end{picture}
}
\end{center}
\caption{Extended Higgs-portal scenario (ESHP): annihilation processes involving particles of the dark sector, $S_i$, $i=1,2$. 
}
\label{fig:EHPprocesses}
\end{figure}
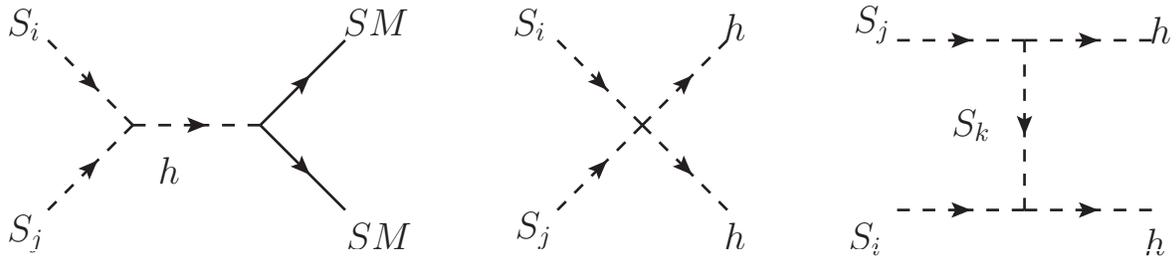

\begin{figure}[t!]
\begin{center}
\fcolorbox{white}{white}{
  \begin{picture}(420,99) (24,-9)
    \SetWidth{1.0}
    \SetColor{Black}
    \Line[dash,dashsize=5,arrow,arrowpos=0.5,arrowlength=5,arrowwidth=2,arrowinset=0.2](48,67)(78,35)
    \Line[dash,dashsize=5,arrow,arrowpos=0.5,arrowlength=5,arrowwidth=2,arrowinset=0.2](108,67)(78,35)
    \Line[dash,dashsize=5,arrow,arrowpos=0.5,arrowlength=5,arrowwidth=2,arrowinset=0.2](78,35)(78,3)
    \Text(45,73)[lb]{\Large{\Black{$S_1$}}}
    \Text(104,73)[lb]{\Large{\Black{$S_1$}}}
    \Text(76,-12)[lb]{\Large{\Black{$h$}}}

    \Line[dash,dashsize=5,arrow,arrowpos=0.5,arrowlength=5,arrowwidth=2,arrowinset=0.2](145,67)(175,53)
    \Line[dash,dashsize=5,arrow,arrowpos=0.5,arrowlength=5,arrowwidth=2,arrowinset=0.2](205,67)(175,53)
    \Line[dash,dashsize=5,arrow,arrowpos=0.5,arrowlength=5,arrowwidth=2,arrowinset=0.2](175,25)(175,3)
    \Arc[dash,dashsize=5,arrow,arrowpos=0.5,arrowlength=5,arrowwidth=2,arrowinset=0.2](175,39)(14,90,270)
    \Arc[dash,dashsize=5,arrow,arrowpos=0.5,arrowlength=5,arrowwidth=2,arrowinset=0.2](175,39)(14,-90,90)
    \Text(142,73)[lb]{\Large{\Black{$S_1$}}}
    \Text(201,73)[lb]{\Large{\Black{$S_1$}}}
    \Text(142,35)[lb]{\Large{\Black{$S_1$}}}
    \Text(196,35)[lb]{\Large{\Black{$S_2$}}}
    \Text(173,-12)[lb]{\Large{\Black{$h$}}}

    \Line[dash,dashsize=5,arrow,arrowpos=0.5,arrowlength=5,arrowwidth=2,arrowinset=0.2](312,67)(295,33)
    \Arc[dash,dashsize=5,arrow,arrowpos=0.5,arrowlength=5,arrowwidth=2,arrowinset=0.2](282,39)(14,135,-45)
    \Arc[dash,dashsize=5,arrow,arrowpos=0.5,arrowlength=5,arrowwidth=2,arrowinset=0.2](282,39)(14,-45,-225)
    \Line[dash,dashsize=5,arrow,arrowpos=0.5,arrowlength=5,arrowwidth=2,arrowinset=0.2](252,67)(271,49)
    \Line[dash,dashsize=5,arrow,arrowpos=0.5,arrowlength=5,arrowwidth=2,arrowinset=0.2](293,30)(293,3)    
    \Text(308,73)[lb]{\Large{\Black{$S_1$}}}
    \Text(249,73)[lb]{\Large{\Black{$S_1$}}}
    \Text(270,58)[lb]{\Large{\Black{$S_2$}}}
    \Text(265,13)[lb]{\Large{\Black{$h$}}}
    \Text(282,-12)[lb]{\Large{\Black{$h$}}}

    \Line[dash,dashsize=5,arrow,arrowpos=0.5,arrowlength=5,arrowwidth=2,arrowinset=0.2](349,67)(365,46)
    \Line[dash,dashsize=5,arrow,arrowpos=0.5,arrowlength=5,arrowwidth=2,arrowinset=0.2](365,46)(394,46)
    \Line[dash,dashsize=5,arrow,arrowpos=0.5,arrowlength=5,arrowwidth=2,arrowinset=0.2](379,20)(365,46)
    \Line[dash,dashsize=5,arrow,arrowpos=0.5,arrowlength=5,arrowwidth=2,arrowinset=0.2](394,46)(379,20)
    \Line[dash,dashsize=5,arrow,arrowpos=0.5,arrowlength=5,arrowwidth=2,arrowinset=0.2](379,20)(379,3)
    \Line[dash,dashsize=5,arrow,arrowpos=0.5,arrowlength=5,arrowwidth=2,arrowinset=0.2](409,67)(394,46)
    \Text(357,32)[lb]{\Large{\Black{$h$}}}
    \Text(378,-12)[lb]{\Large{\Black{$h$}}}
    \Text(395,32)[lb]{\Large{\Black{$h$}}}
    \Text(365,56)[lb]{\Large{\Black{$S_2$}}}
    \Text(346,73)[lb]{\Large{\Black{$S_1$}}}
    \Text(405,73)[lb]{\Large{\Black{$S_1$}}}
  \end{picture}
}
\end{center}
\caption{Tree-level $S_1S_1h$ vertex and main 1-loop corrections.}
\label{fig:1loopvertices}
\end{figure}
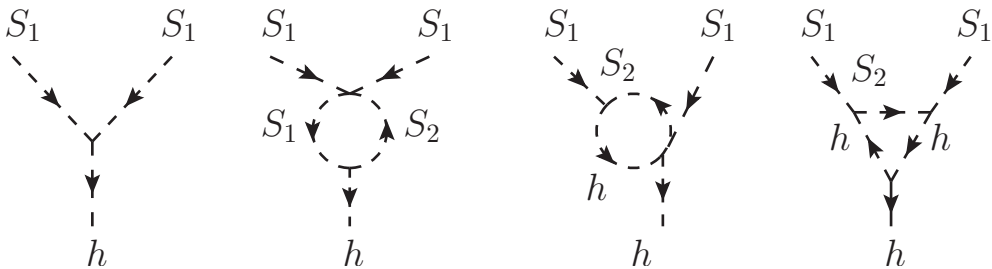

Moreover, all the processes where a $\lambda_1-$vertex is involved get new radiative corrections. In particular, the trilinear vertex (after EW breaking) $S_1S_1h$, which appears in DM annihilation and scattering processes (relevant for indirect and direct detection), has to be corrected by 1-loop diagrams, such as the ones depicted in Fig.~\ref{fig:1loopvertices}. Due to the adopted smallness of $\lambda_2$, other 1-loop diagrams are subdominant. Assuming for simplicity
that $\lambda_{31}$ (involved in the second diagram of Fig.~\ref{fig:1loopvertices}) is of the same order as $\lambda_{12}$, all these contributions are ${\cal O}(\lambda_{12}^2/(4\pi)^2)$, which is precisely the smallest natural value for $\lambda_1$. This means that only  when $\lambda_1$ is close to this lower
limit the contributions of these diagrams may be significant{\color{magenta}\footnote{In that case, there may be accidental cancellations between the tree-level and the radiative corrections, as can be checked from the explicit expressions given in the Appendix. Moreover these cancellations can be more or less significant depending on the external momenta entering the vertex. This opens the possibility of blind spots for direct or indirect detection, while keeping a sizable annihilation in the early universe.}}. 
Nevertheless, for consistency, we have included the contribution of the 1-loop diagrams in all cases. A detailed discussion of these radiative corrections is given in the Appendix.

Let us now turn our attention to the computation of the relic density. We will start by considering a scenario in which $\lambda_1$ is as small as possible ($\lambda_1 = \lambda_{12}^2/(4\pi)^2$). Then, $\lambda_1$ can be neglected for all the relevant physical processes in most cases, so the only significant parameters to describe the DM physics are $\msl$, $\msh$, and $\lambda_{12}$.
For each  value of the DM mass, $\msl$, we are interested in finding out which combinations of $\msh$ and $\lambda_{12}$ lead to the correct relic density. 

\begin{figure}[t!]
\centering 
\hspace*{-1ex}
\includegraphics[width=0.33\linewidth]{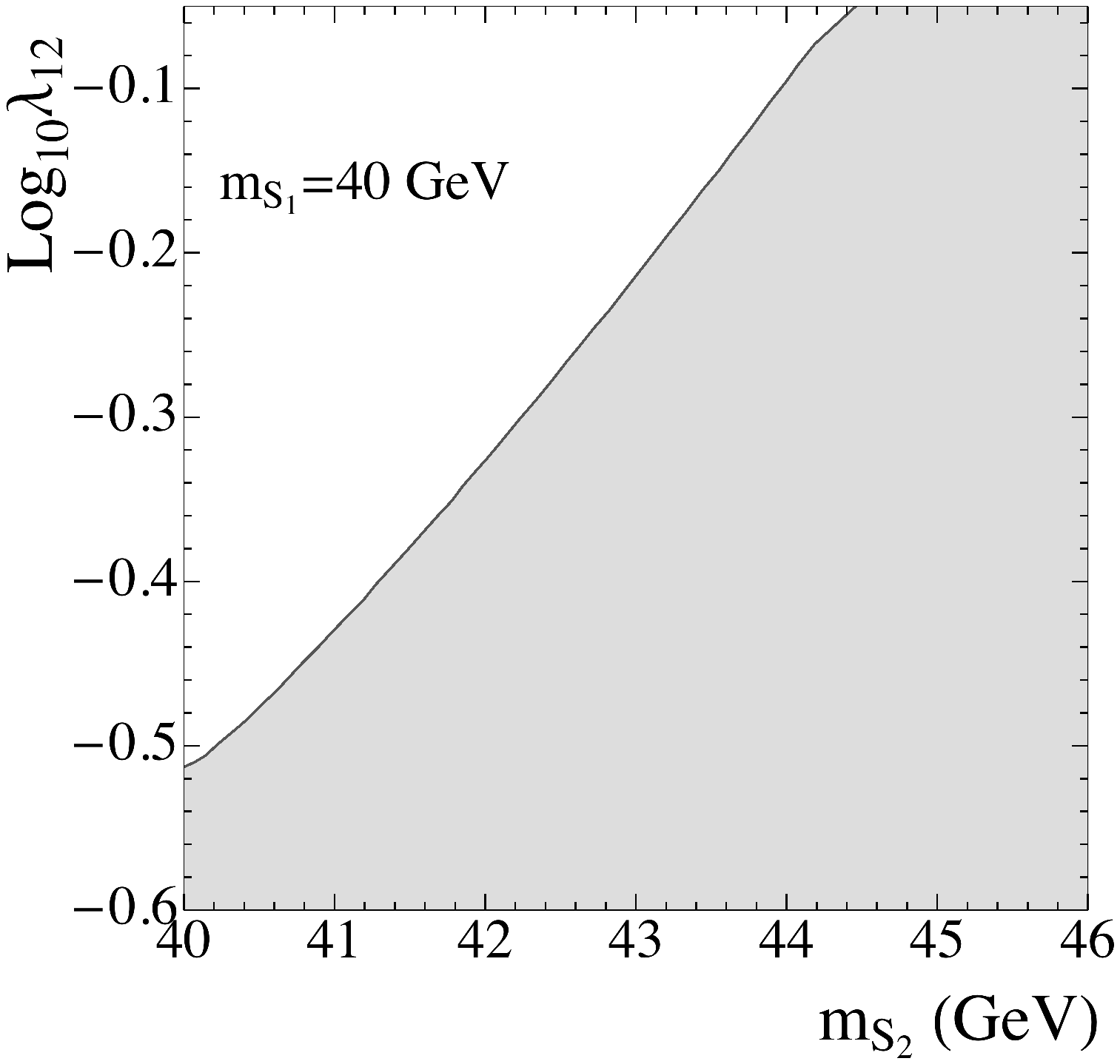}
\hspace*{-1ex}
\includegraphics[width=0.33\linewidth]{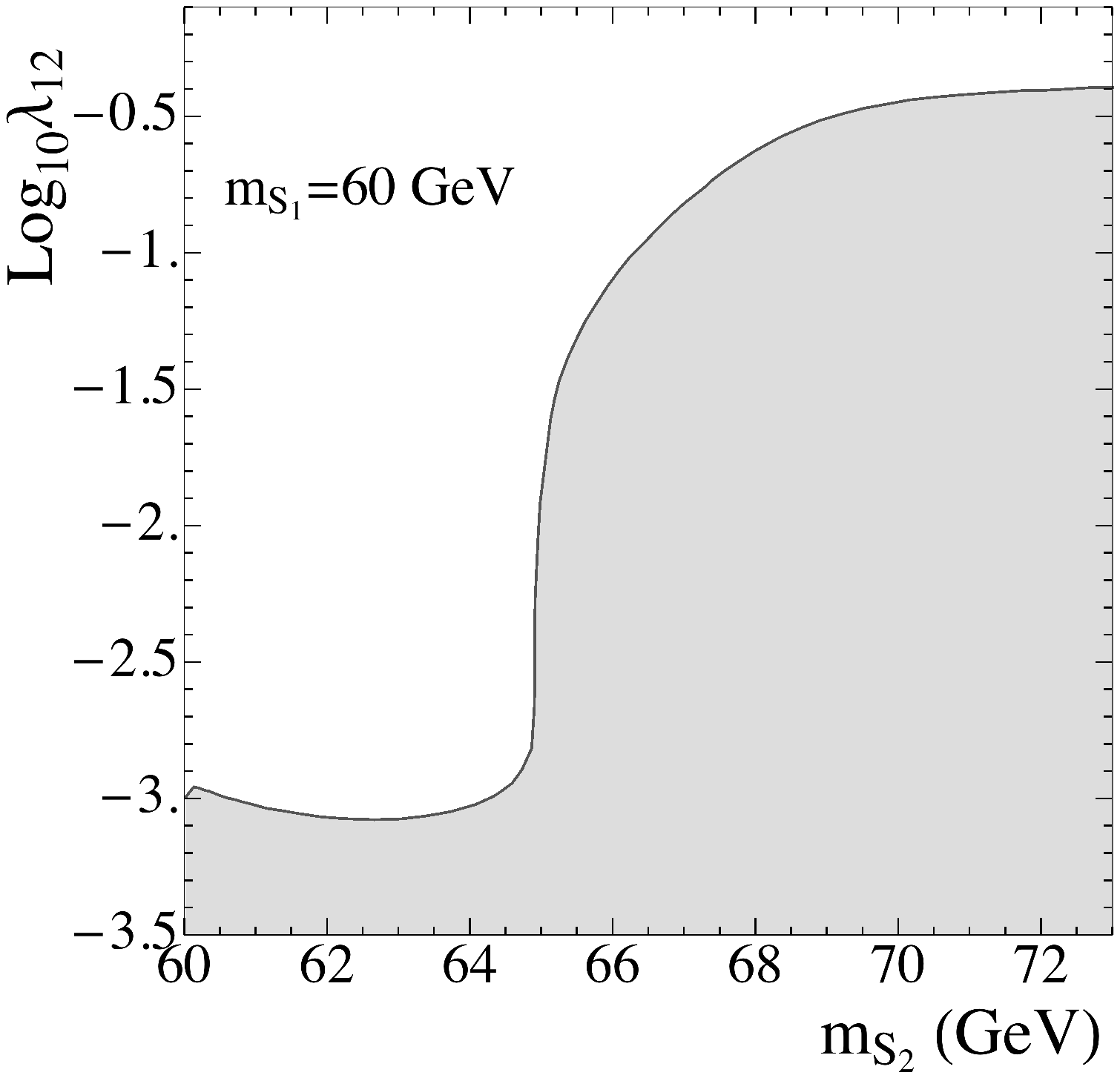}
\hspace*{-1ex}
\includegraphics[width=0.33\linewidth]{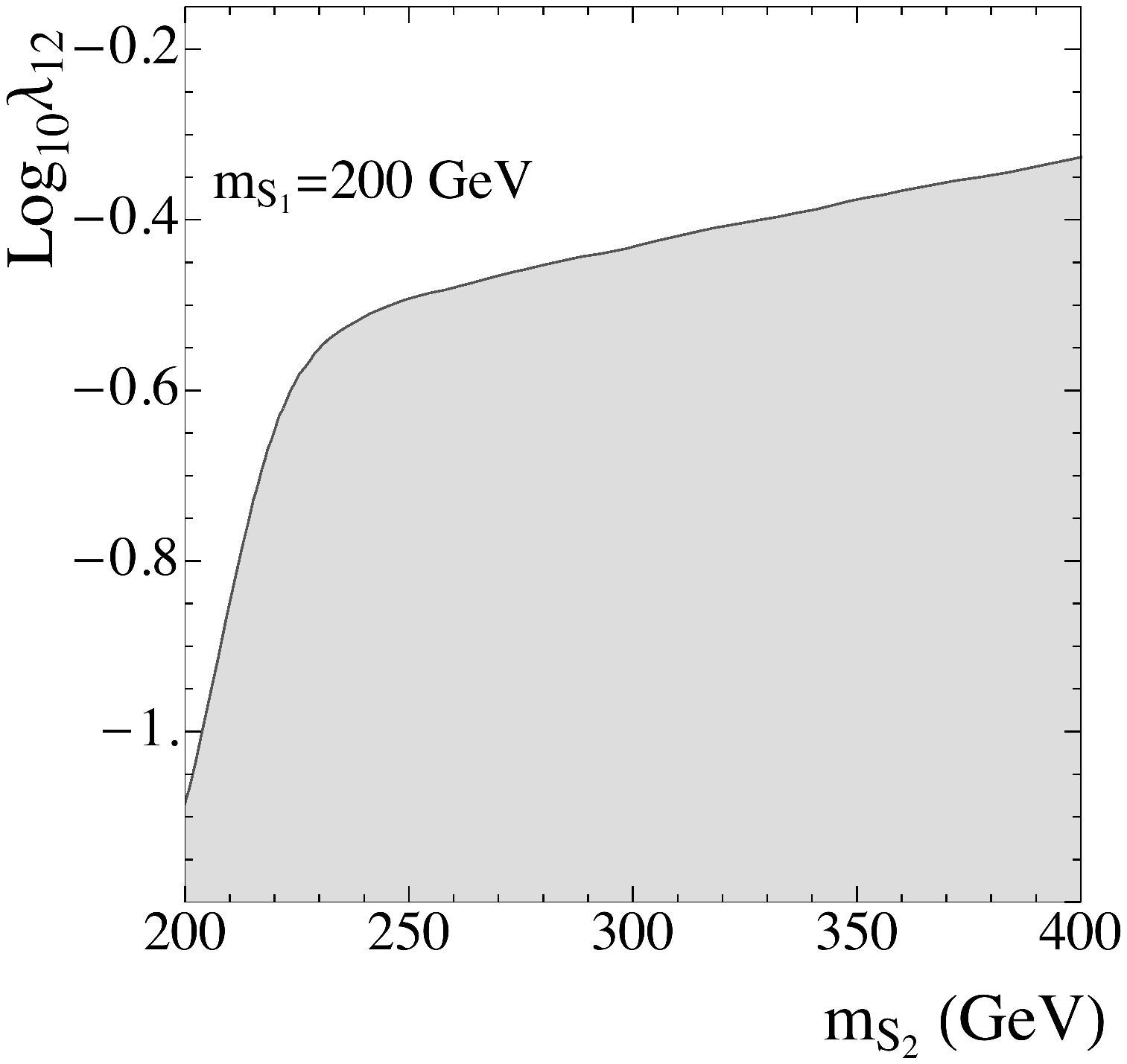}
\caption{
Range of values in the $\left\{\lambda_{12},\,m_{S_2}\right\}$ plane leading to the correct DM relic density
for three illustrative values of the DM mass: (from left to right) $\msl=40$~GeV, 60~GeV, and 200~GeV. The DM-Higgs coupling has been fixed to $\lambda_{1}=\lambda_{12}^2/(4\pi)^2$.
The solid black line represents the Planck result. The grey area below this line is excluded since $\Omega_{S_1}>\Omega_{CDM}$.
}
\label{fig:3examples}
\end{figure}

Fig.\,\ref{fig:3examples} shows the line along which the correct DM relic abundance is obtained 
for three representative cases, namely $\msl=40$, $60$, and $200$~GeV, i.e., below, around and above the Higgs resonance (left, middle and right panels, respectively). Let us discuss each case separately.

For small DM masses (left panel),
the correct relic density can be obtained through coannihilation effects with $S_2$ for a wide range of values of $\lambda_{12}$ when $\msh-\msl\lesssim 5$~GeV.
As $\msh$ grows and departs from $\msl$, the required value of $\lambda_{12}$ is larger and, at some point, it becomes non-perturbative.

When $\msl$ is not far from the Higgs resonance (middle panel), we observe two different regimes. If $\msl+\msh$ is smaller than $m_h$, but such that $\msl+\msh\approx m_h$,
the resonant condition for the $s$-channel $S_1S_2\rightarrow h \rightarrow SM\ SM$ can still be satisfied ($S_1$ and $S_2$ can have the correct energy due to their kinetic energy in the thermal bath) and the required value of $\lambda_{12}$ is small. 
\footnote{Actually, it is quite 
independent of $\msh$, for the following reason. The amount of DM annihilated in this way is proportional to the product of two Boltzman factors: the one that suppresses the $S_2-$density and the one that kinematically suppresses the $S_1S_2\rightarrow h$ process. As $\msh$ increases, the first Boltzman factor decreases and the second one increases, keeping the product almost constant.}
On the other hand, when $\msl+\msh>m_h$ the resonant effect is not possible. Consequently
$\lambda_{12}$ has to increase to reproduce the correct relic density. 
For sufficiently large $\msh$ and $\lambda_{12}$, the corresponding value of the $\lambda_1$ coupling (which in this example is set to $\lambda_1=(\lambda_{12}/4\pi)^2$) and the size of the 1-loop diagrams of Fig.~\ref{fig:1loopvertices} become large enough for the DM to be efficiently annihilated through the usual SHP process, $S_1 S_1\rightarrow h\rightarrow {\rm SM\ SM}$. In this regime, the model works essentially as the conventional SHP and the $S_2$ particle is irrelevant. Then the line in the plot becomes horizontal since the required value of $\lambda_1$ is related to that of $\lambda_{12}$ through the above identification. However, the model could also work with essentially the same $\lambda_1$ and a smaller $\lambda_{12}$.

Finally, for $\msl>m_h/2$ (right panel), we can distinguish two regimes. When $\msh\sim \msl$, coannihilation effects are still present and the dependence with $\lambda_{12}$ resembles that of the left panel. However, for large $\msh$ coannihilation effects are not effective and the relic density becomes less sensitive to $\msh$. In that case, if $\msl>m_h$ (as in the example of the figure), the $t$-channel diagram of Fig.~\ref{fig:3examples}, with $S_1$ in the external legs annihilating through $S_2-$exchange into a pair of Higgs bosons, is kinematically accessible and it becomes the main contribution to the annihilation cross section.

\subsection{Observational and experimental constraints}
\label{sec:constraints}

From the discussion in the previous subsection, it seems that for any value of $\msl$, we can suitably choose $\{\msh,\, \lambda_{12}\}$ to reproduce the correct relic density. Since $\lambda_1$ can be very small, one might expect that the ESHP model can evade easily the usual constraints on the singlet-scalar Higgs-portal.

However, this is not so straightforward. 
First, a sizable $\lambda_{12}$ has potential impact on several observables, as we are about to see. Also, one must check that the existence of the second dark (unstable) species, $S_2$, does not produce any cosmological disaster in the early universe. Finally, we might actually be interested in varying the value of $\lambda_1$ above its minimal value (in order to be as general as possible).

In this subsection we discuss the various physical constraints to which the model is subject.

\paragraph{Invisible width of the SM Higgs boson.}
From the observed decay channels of the SM Higgs boson, an experimental constraint can be derived on its invisible decay width. Namely, using the recent ATLAS and CMS results \cite{Aad:2015txa,Khachatryan:2016whc,Khachatryan:2016vau,CMS:2016rfr}, we will impose BR$(h\to inv)\le 0.20$ (at a $90\%$ confidence level) throughout this article. 
In the scenario presented here, the DM sector can contribute to the invisible width of the SM Higgs through the decays $h\to S_1S_1$, $h\to S_1S_2$, and $h\to S_2S_2$, when these are kinematically allowed (see also Ref.\,\cite{Ghorbani:2014gka}).

The corresponding decay widths at tree level read 
\begin{eqnarray}
\Gamma_{h\to S_1S_1}&=&\frac{\lambda_1^2 v^2}{32\pi m_h}
\left(
1-\frac{4\,\msl^2}{m_h^2}
\right)^{1/2},\nonumber\\
\Gamma_{h\to S_1S_2}&=&\frac{\lambda_{12}^2 v^2}{64\pi m_h}
\left(
1-\frac{(\msh+\msl)^2}{m_h^2}
\right)^{1/2}
\left(
1-\frac{(\msh-\msl)^2}{m_h^2}
\right)^{1/2}
,\nonumber\\
\Gamma_{h\to S_2S_2}&=&\frac{\lambda_2^2 v^2}{32\pi m_h}
\left(
1-\frac{4\,\msh^2}{m_h^2}
\right)^{1/2}.
\label{eq:inv}
\end{eqnarray}
In our calculation, we have included the radiative corrections to the $S_1S_1h$ coupling (see Fig.\ref{fig:1loopvertices}), as explained in the previous section.
As mentioned in the Introduction, in the conventional SHP this constraint excludes areas with large coupling for small dark matter masses. 
In the ESHP, both $\lambda_1$ and $\lambda_2$ can be chosen small and, therefore, $h\to S_1S_2$ is the most relevant process, setting an upper bound on $\lambda_{12}$.

\paragraph{Lifetime of the extra scalar particle.}
The heavy scalar $S_2$ is unstable and decays into $S_1$ (plus SM products). We will require that the decay occurs before Big Bang nucleosynthesis, so as not to spoil its predictions. In fact, if $S_2$ is substantially heavier than $S_1$, namely $\msh>\msl+m_h$, it rapidly decays as $S_2\to S_1 h$ through the corresponding trilinear coupling $\lambda_{12}$. However, if $\msh<\msl+m_h$, we need to consider the three-body decay $S_2\to S_1 f\bar f$. The latter is in general fast enough when the $S_1 b\bar b$ channel is open, but the lifetime of $S_2$ increases rapidly below this mass.
We have computed the lifetime of $S_2$ using CalcHEP \cite{Belyaev:2012qa}, and excluded points in the scan where $\tau_{S_2}>1$~s.

\begin{figure}[t!]
\begin{center}
\fcolorbox{white}{white}{
  \begin{picture}(130,99) (159,-9)
    \SetWidth{1.0}
    \SetColor{Black}
    \Line[dash,dashsize=5,arrow,arrowpos=0.5,arrowlength=5,arrowwidth=2,arrowinset=0.2](160,70)(224,50)
    \Line[dash,dashsize=5,arrow,arrowpos=0.5,arrowlength=5,arrowwidth=2,arrowinset=0.2](224,50)(224,5)
    \Line[arrow,arrowpos=0.5,arrowlength=5,arrowwidth=2,arrowinset=0.2](224,8)(288,-9)
    \Line[arrow,arrowpos=0.5,arrowlength=5,arrowwidth=2,arrowinset=0.2](160,-9)(224,8)
    \Line[dash,dashsize=5,arrow,arrowpos=0.5,arrowlength=5,arrowwidth=2,arrowinset=0.2](224,50)(288,70)
    \SetColor{Gray}
    \Vertex(224,50){10}
    \Text(155,73)[lb]{\Large{\Black{$S_1$}}}
    \Text(287,73)[lb]{\Large{\Black{$S_1$}}}
    \Text(155,-22)[lb]{\Large{\Black{$q$}}}
    \Text(287,-22)[lb]{\Large{\Black{$q$}}}
    \Text(210,20)[lb]{\Large{\Black{$h$}}}
  \end{picture}
}
\end{center}\caption{Diagrams contributing to the direct detection of $S_1$. The gray circle represents the sum of the (tree-level and 1-loop) vertices of Fig.~\ref{fig:1loopvertices}.
} 
\label{fig:dd-penguin}
\end{figure}
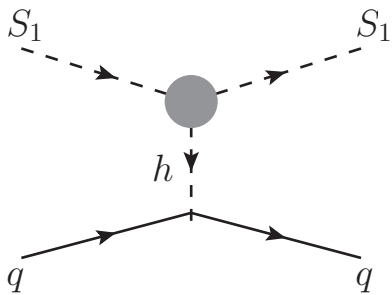

\paragraph{Direct detection.}
\hspace{0cm}
The tree-level scattering of $S_1$ off quarks occurs via a $t$-channel Higgs exchange, as depicted in Fig.\,\ref{fig:dd-penguin}, where the gray circle represents the sum of the (tree-level and 1-loop) vertices of Fig.~\ref{fig:1loopvertices}. Since $\lambda_1$ can be very small,
the constraints from direct detection experiments are substantially alleviated, in contrast with the situation of the canonical Higgs portal, as has also been observed in Ref.\,\cite{Ghorbani:2014gka}.

We have explicitly computed the spin-independent contribution to the DM-nucleon elastic scattering cross section, $\sigsi$, which occurs through the exchange of a Higgs boson, as illustrated in Fig.\,\ref{fig:dd-penguin}. The Higgs-nucleon coupling can be parametrized as $f_N m_N /v$ where $m_N\simeq 0.946$ GeV is the mass of the nucleon.   According to this, the spin-independent cross section, $\sigsi$, reads
\begin{equation}
\sigsi
\label{sics}=
\frac{\lambda_1^2 f_N^2 \mu^2 m_N^2}{4 \pi m_h^4 m_{S_1}^2}\ ,
\end{equation}
where $\mu=m_N m_{S_1}/(m_N+m_{S_1})$ is the nucleon-DM reduced mass. The $f_N$ parameter contains the nucleon matrix elements,
and its full expression can be found, e.g., in ref.~\cite{Cline:2013gha}. 
Using the values for the latter obtained from the lattice evaluation
\cite{Alarcon:2011zs,Alarcon:2012nr,Alvarez-Ruso:2013fza,Young:2013nn,Abdel-Rehim:2016won,Duerr:2016tmh}, one arrives
at $f_N=0.30 \pm 0.03$, in agreement with Ref.\,\cite{Cline:2013gha}. 
Finally,
we have included one-loop contributions to the $S_1S_1h$ coupling, shown in Fig.\,\ref{fig:1loopvertices}, according to the computation given in the Appendix.

Then, we have implemented the most recent upper bounds obtained by the LUX collaboration \cite{Akerib:2016vxi} (which improves the bound obtained by PandaX-II \cite{Tan:2016zwf}) for DM particles with masses above 6~GeV.\footnote{The SuperCDMS \cite{Agnese:2014aze} and CRESST \cite{Angloher:2015ewa} collaborations have obtained more stringent constraints for light DM particles, but this range of masses is excluded in our model, mainly because of the experimental constraint on the invisible decay width of the Higgs boson.} Notice that since $S_1$ is a scalar field, there is no contribution to spin-dependent terms.

Although in principle we could also have inelastic scattering processes at tree level, $S_1 q\to S_2 q$, the typical mass difference in our scenario is such that $\msh-\msl> 1$~GeV, significantly larger than the kinetic energy of the incoming DM particle (which is smaller than $\sim1$~MeV for DM particles lighter than $\sim1$~TeV), and this process does not take place.

\paragraph{Indirect detection}
\hspace{0cm}
Regarding indirect dark matter searches, the most relevant bounds for this model can be derived from gamma-ray searches from dwarf spheroidal galaxies (for the continuum spectrum) and the galactic centre (for gamma ray lines and spectral features).

In order to apply the dwarf spheroidal galaxies data on the continuum, we have computed the thermally-averaged annihilation cross section, $\langle \sigma v \rangle$, in the dwarf galaxies using MicrOMEGAs \cite{Belanger:2013oya,Belanger:2014vza}, assuming that the initial particles are at rest (a good approximation since the velocity of the DM is low). 
We have then confronted the results with the combined analysis of Fermi-LAT and Magic \cite{Ahnen:2016qkx}, considering the upper bounds on $\langle \sigma v \rangle$ for annihilation into $b\bar b$ (again a good approximation since the annihilation is through the Higgs and this is the main final state when it is open).

On the other hand, for gamma ray lines in the galactic centre, we have calculated the annihilation cross section into a pair of photons, $\langle \sigma v \rangle_{\gamma \gamma}$, again using MicrOMEGAs, and confronted it with the upper bound given by Fermi-LAT  \cite{Ackermann:2015lka}. We have chosen the Einasto \cite{Einasto1969,Navarro:2003ew} profile for the DM halo, since is more restrictive than Navarro-Frenk-White (NFW) \cite{Navarro:1995iw,Navarro:1996gj} and has a good fit to results of numerical simulations. 
As in the SHP model, a Breit-Wigner
enhancement near the Higgs resonance takes place,\footnote{This has been studied in various models \cite{Feldman:2008xs,Ibe:2008ye,Guo:2009aj,AlbornozVasquez:2011js,Chatterjee:2014bva}.}  although,
given the small decay width of the Higgs boson, it only occurs for a narrow range of masses.
This leads to a sizable annihilation cross section in that region.

Finally, let us recall that indirect detection constraints are very sensitive to whether the gamma-ray flux is re-scaled by the dark matter density squared ($\xi^2$).

\section{Results}
\label{sec:results}

In this section we explore the parameter space of the ESHP model, incorporating all the experimental constraints and computing the theoretical predictions of observables for direct and indirect DM searches.
As mentioned in the previous section, we have used MicrOMEGAs \cite{Belanger:2013oya} to compute the relic abundance and indirect detection observables (the thermal average of the annihilation cross section of $S_1$ particles in the DM halo, $\sigmavh$, and the resulting gamma-ray flux). The spin independent $S_1$-nucleon scattering cross section, $\sigsi$, and the invisible Higgs decay width, have been computed including one-loop corrections, as explained in Section\,\ref{sec:constraints}.

In order to facilitate the comparison of the model with the usual SHP, we have carried out a series of numerical scans, for fixed values of $\lambda_{12}$, in the three dimensional parameter space $\{\msl,\,\lambda_1,\,\msh\}$, searching for points where $S_1$ is a viable candidate for dark matter. Note that the first two parameters are those of the SHP, i.e. the mass and quartic coupling of the DM.
As already mentioned, we will set $\lambda_2$ at its lowest natural value, $\lambda_2=\lambda_{12}^2/(4\pi)^2$. This is also the lower limit of $\lambda_1$ in the scans.

\begin{figure}[t!]
\centering
\setlength{\tabcolsep}{0pt} 
\renewcommand{\arraystretch}{0} 
\begin{tabular}{c c c c}
  & $\Gamma_H^{inv} $ & $\Gamma_H^{inv}+ID $ & $\Gamma_H^{inv}+ID+DD $ \\ 
\begin{sideways} $\qquad \qquad \qquad \lambda_{12}=$0.01 \end{sideways}  $\quad$& \includegraphics[width=0.32\linewidth]{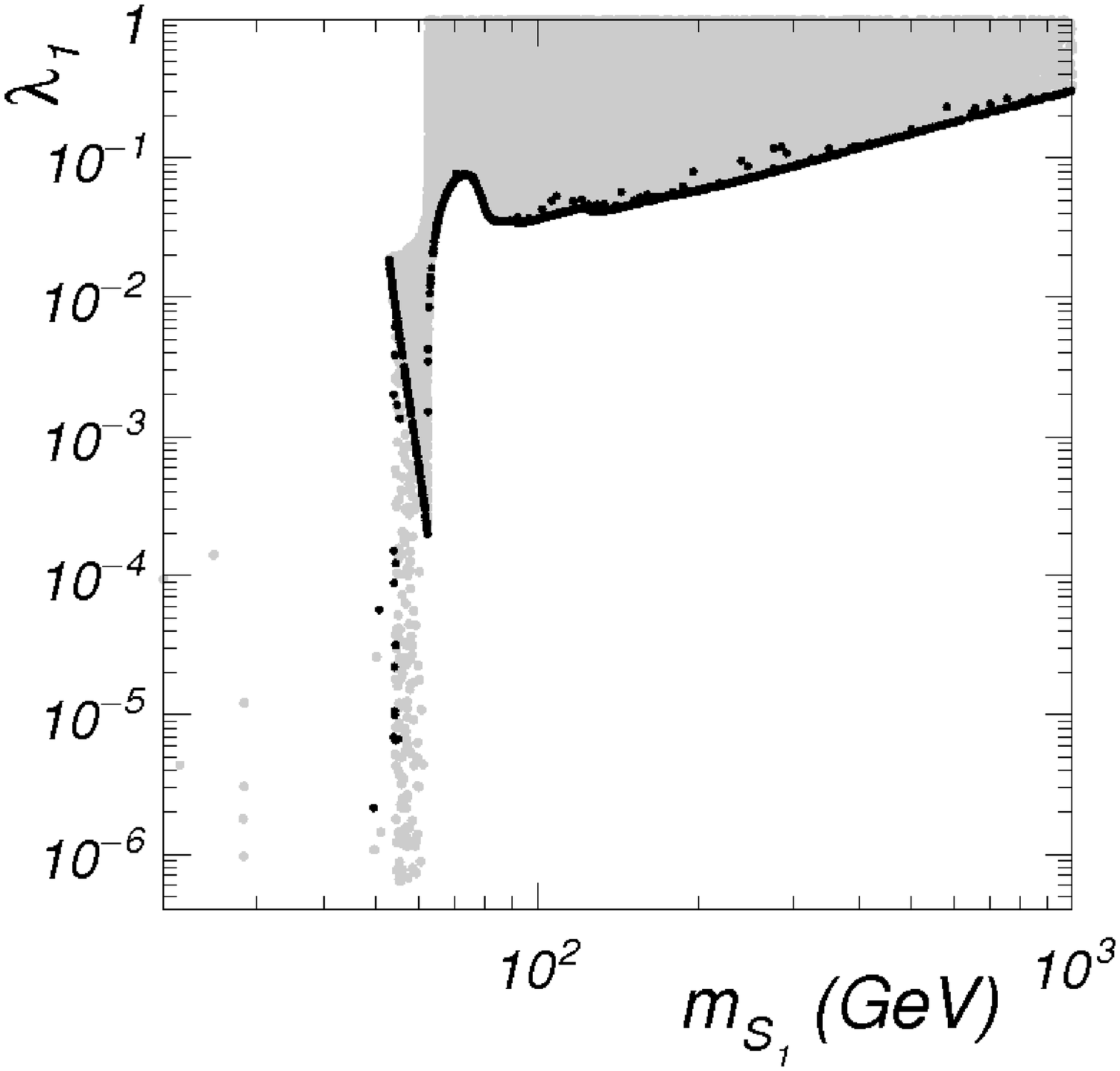} & \includegraphics[width=0.32\linewidth]{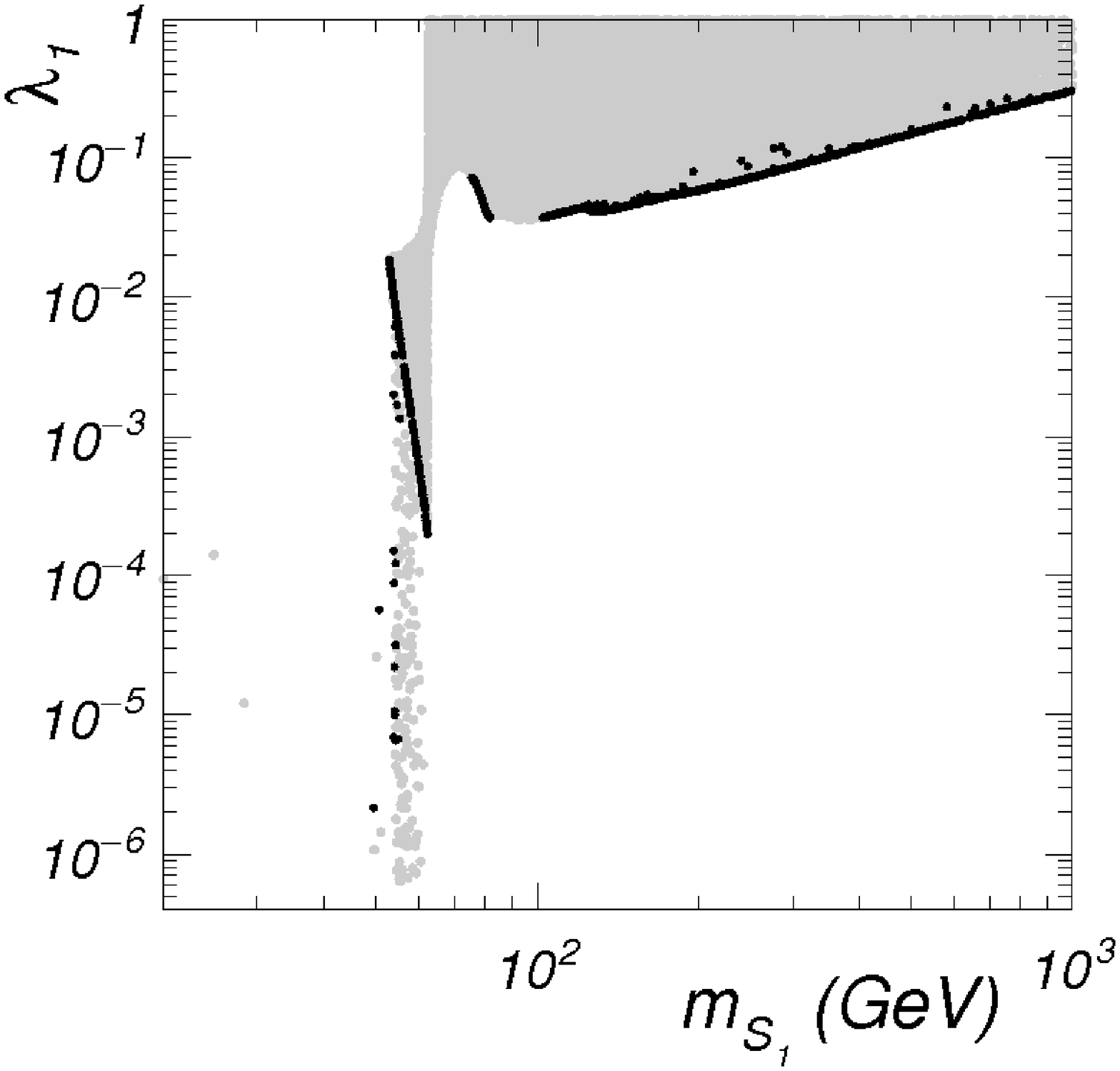} & \includegraphics[width=0.32\linewidth]{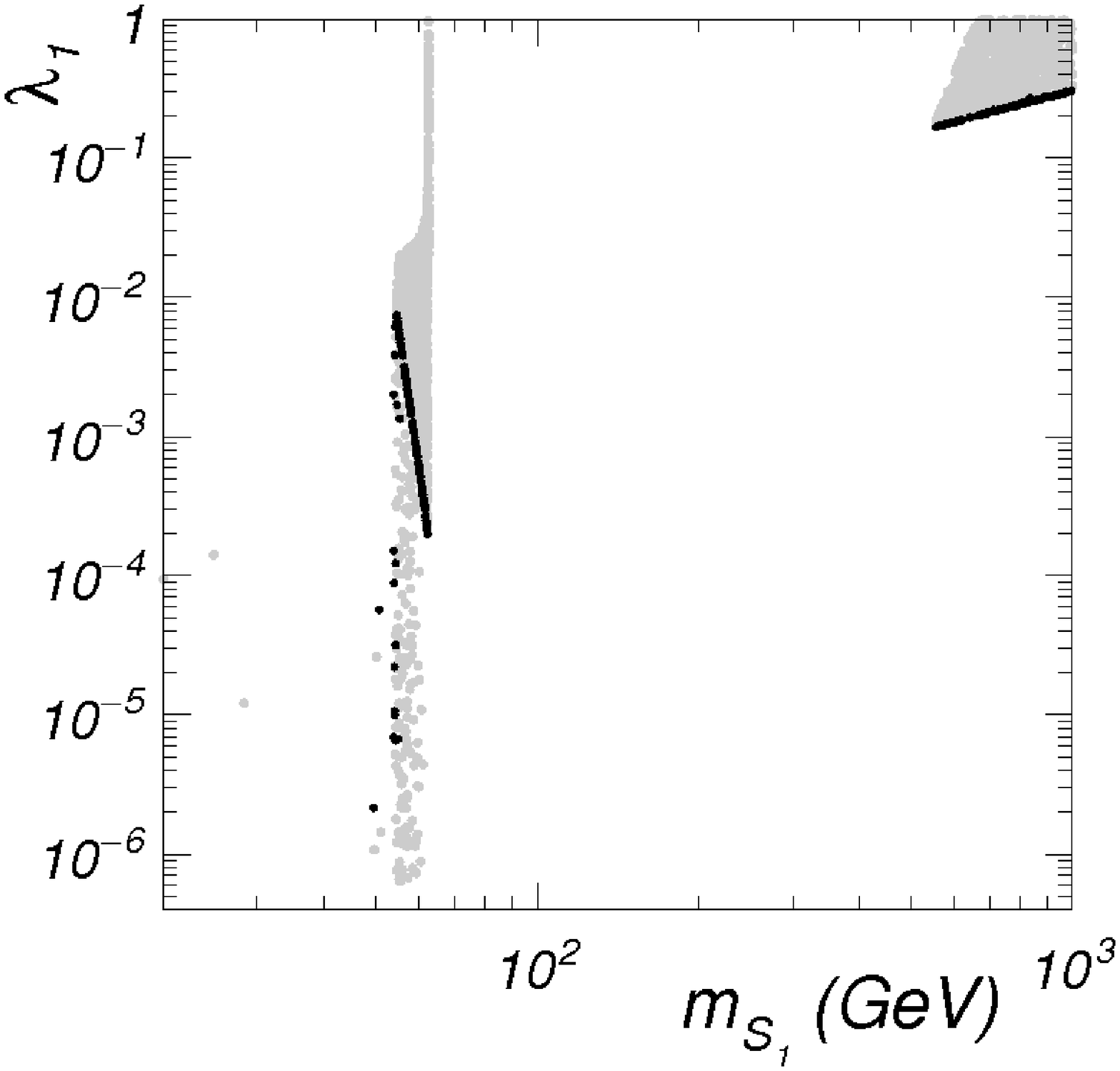} \\ [1ex]
\begin{sideways}$\qquad \qquad \qquad \lambda_{12}=$0.1  \end{sideways}& \includegraphics[width=0.32\linewidth]{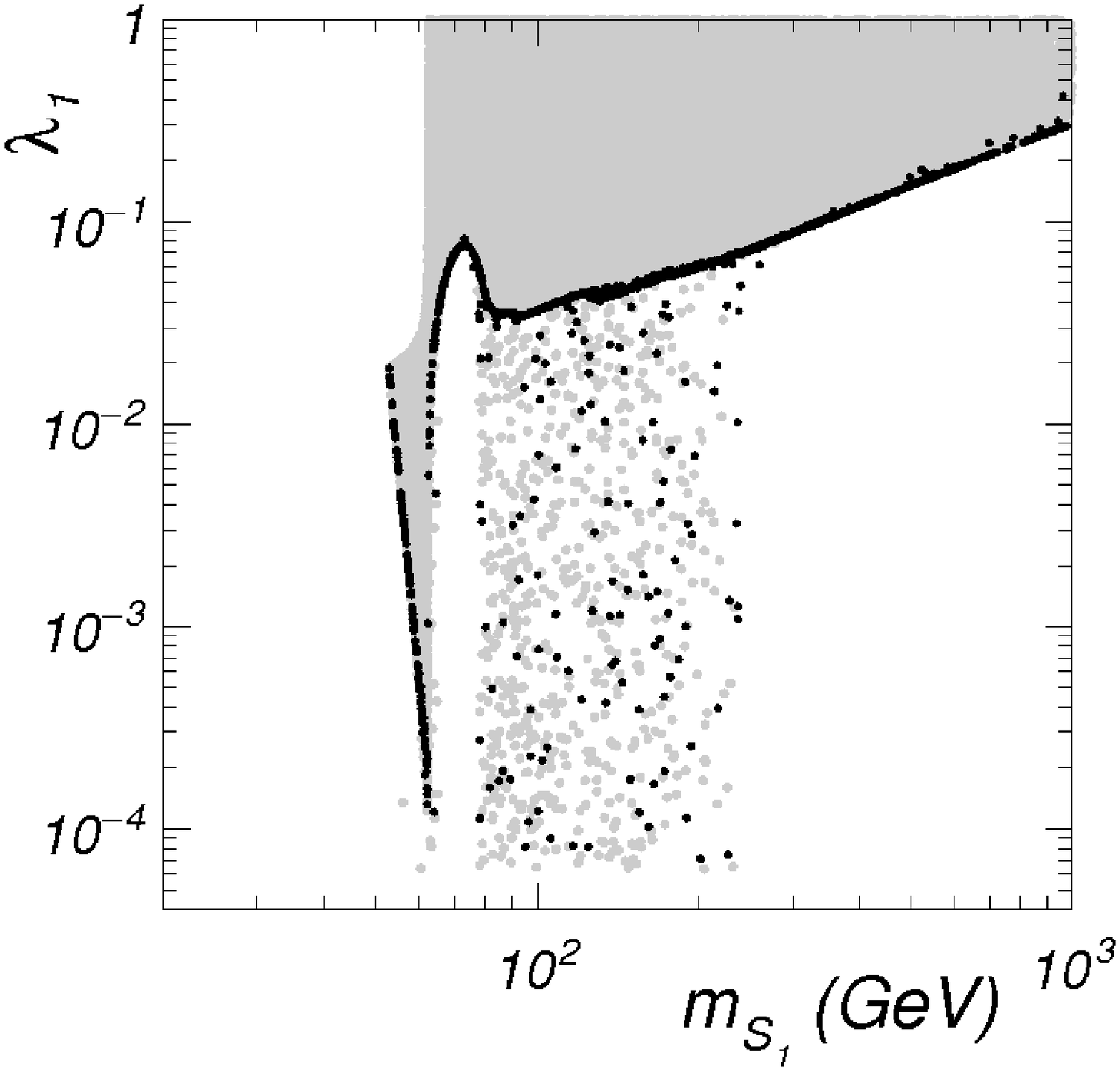} & \includegraphics[width=0.32\linewidth]{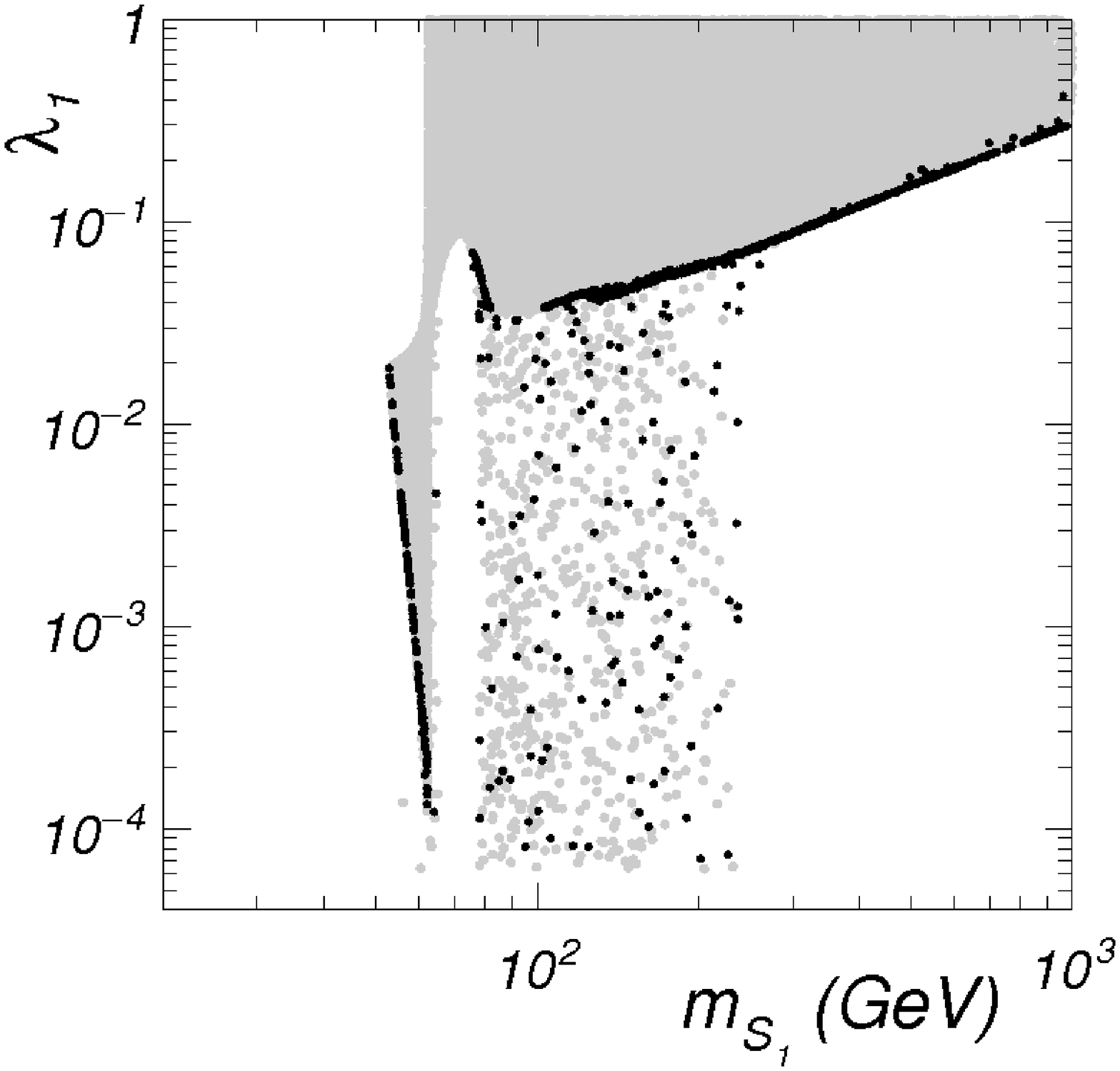} & \includegraphics[width=0.32\linewidth]{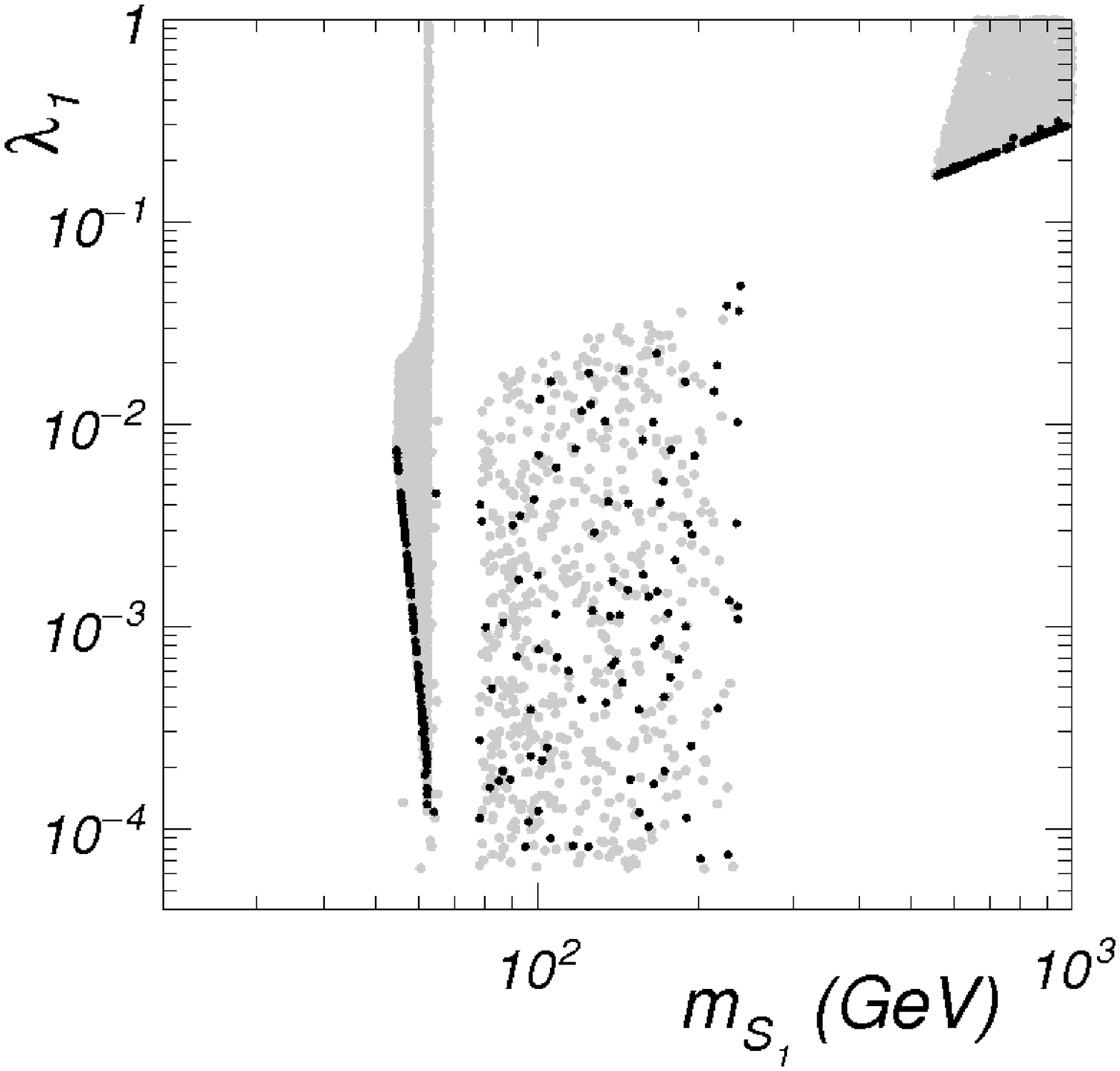} \\ 
\begin{sideways} $\qquad \qquad \qquad \lambda_{12}=1$  \end{sideways}& \includegraphics[width=0.32\linewidth]{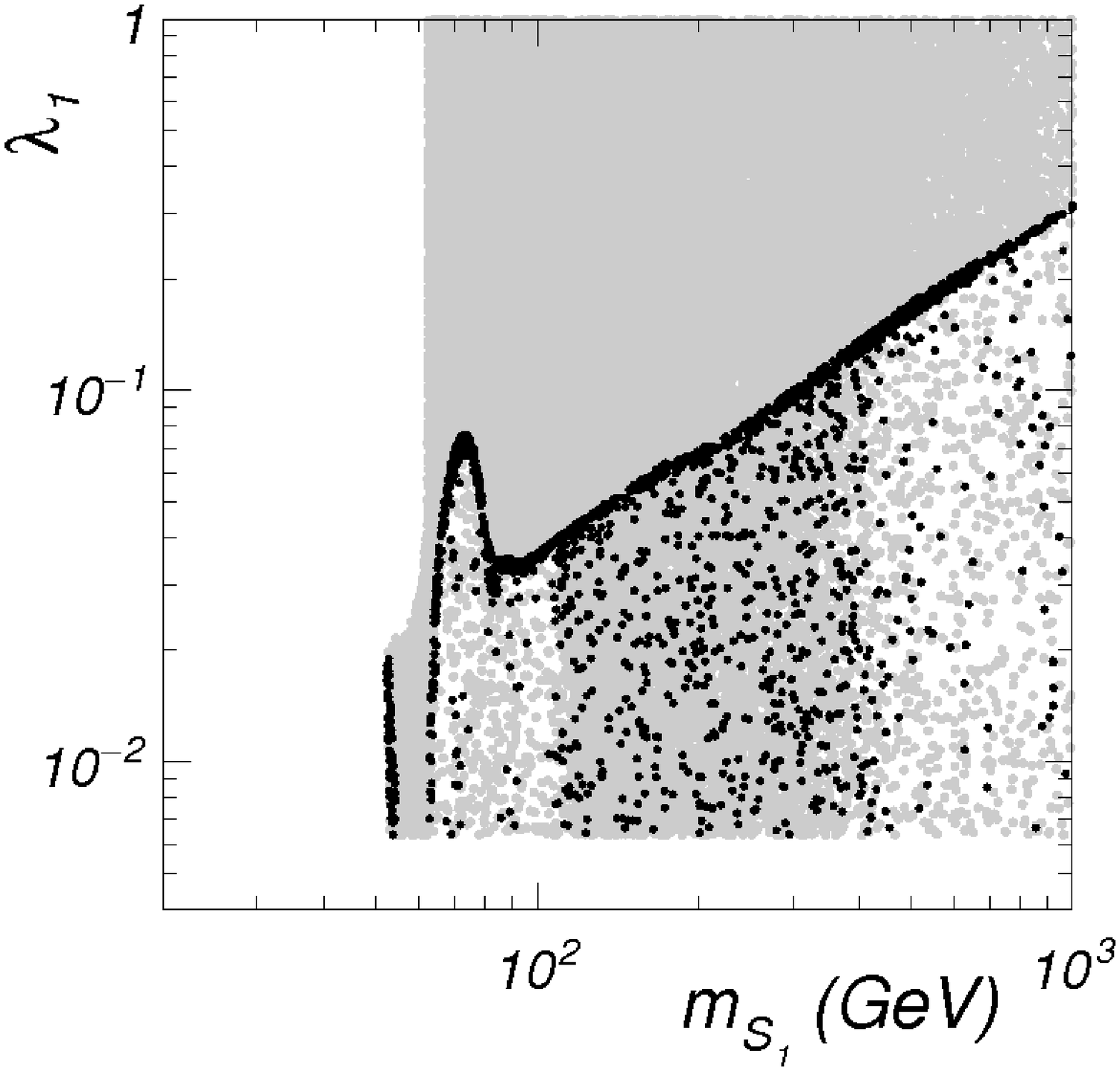} & \includegraphics[width=0.32\linewidth]{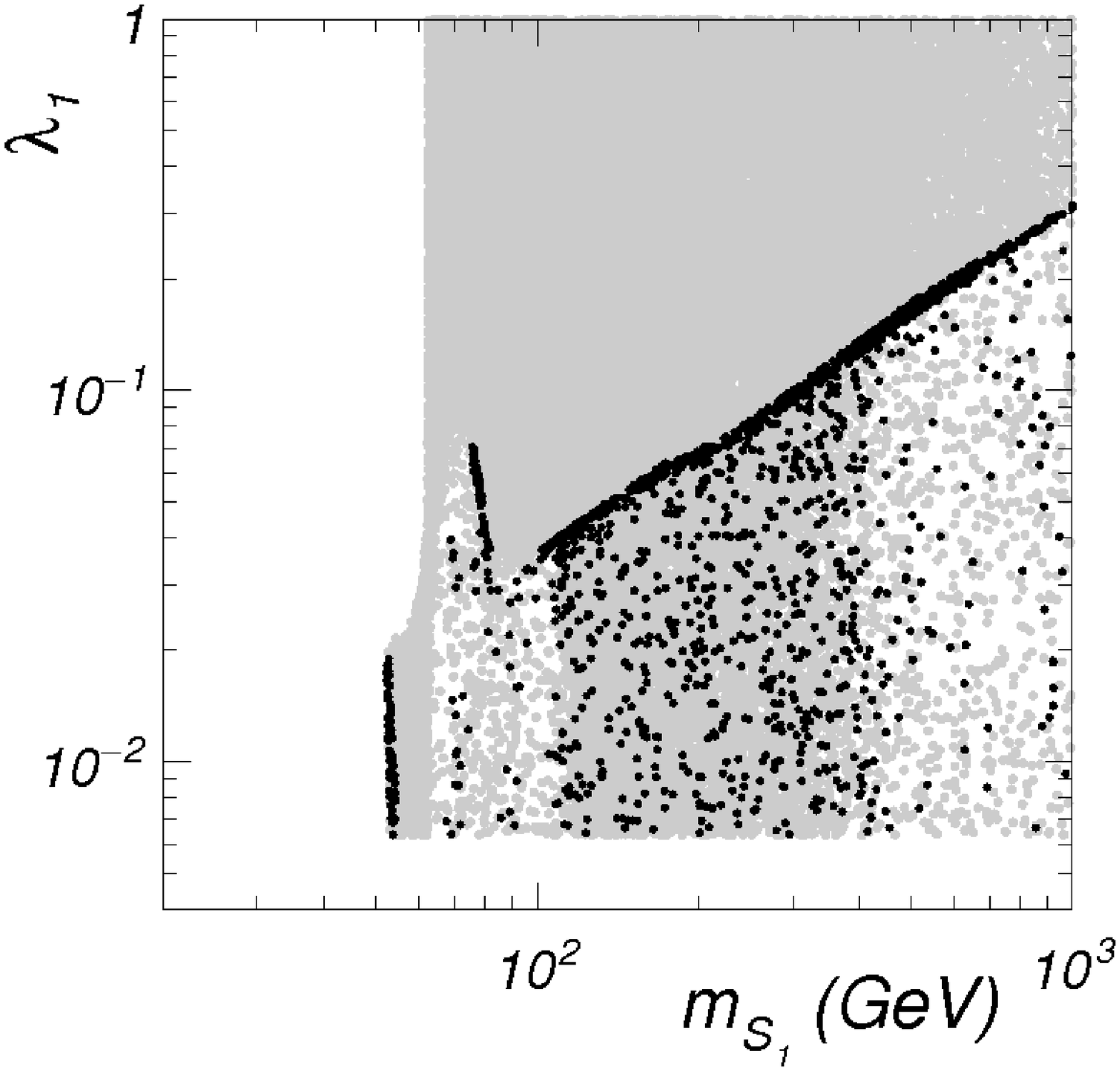} & \includegraphics[width=0.32\linewidth]{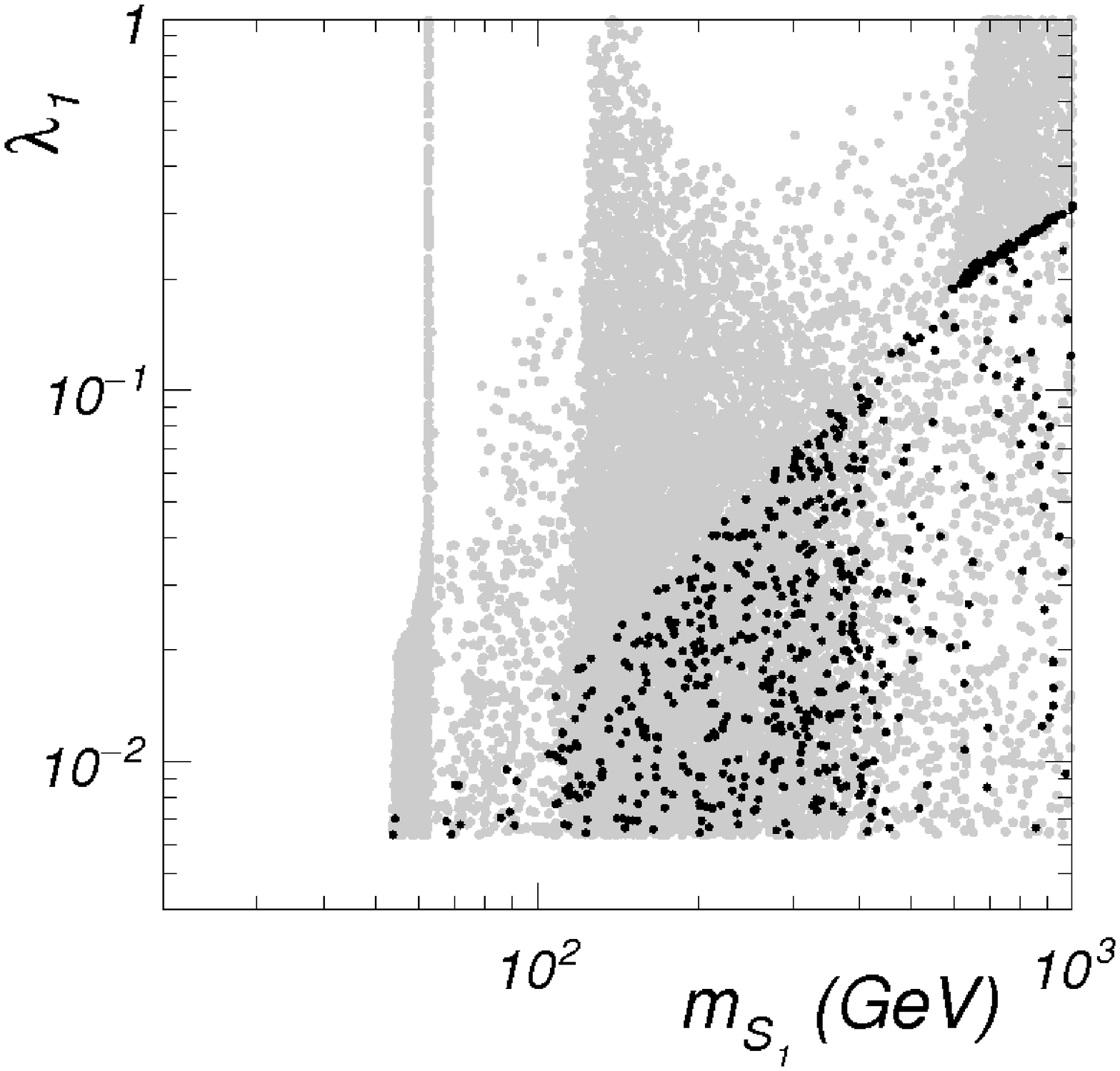} \\ 
\end{tabular}
\caption{
Effect of the experimental constraints in the $\{\lambda_1,\,\msl\}$ parameter space of the ESHP model. 
From up to down, we have fixed $\lambda_{12}=0.01,\,0.1,\,1$, and $\lambda_2=\lambda_{12}^2/(4\pi)^2$.
In all the plots, black (gray) points correspond to those where $\Omega h^2=0.119\pm 0.003$ ($\Omega h^2<0.116$). 
The left column incorporates only constraints from lifetime of $S_2$ and invisible decay width of the Higgs boson. The central column includes also the indirect detection (dSph and gamma ray lines). Finally, the bottom row includes the bound from the LUX constraint.
}
\label{fig:l1m1}
\end{figure}

\begin{figure}[t!]
\centering
\setlength{\tabcolsep}{0pt} 
\renewcommand{\arraystretch}{0} 
\begin{tabular}{c c c c}
  & $\Gamma_H^{inv} $ & $\Gamma_H^{inv}+ID $  & $\Gamma_H^{inv}+ID+DD $ \\ 
\begin{sideways} $\qquad \qquad \qquad \lambda_{12}=$0.01 \end{sideways} $\quad$ & \includegraphics[width=0.32\linewidth]{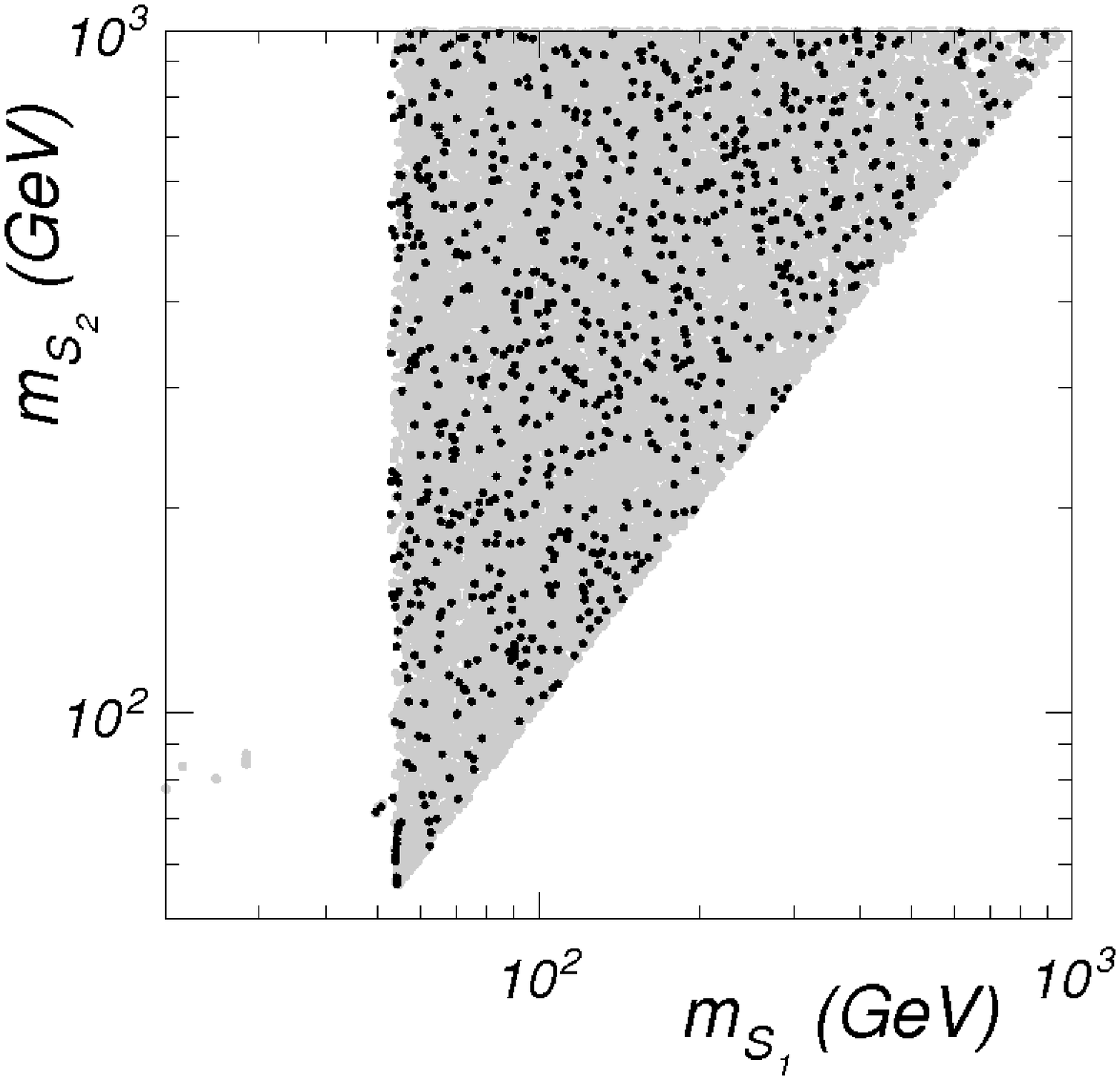} & \includegraphics[width=0.32\linewidth]{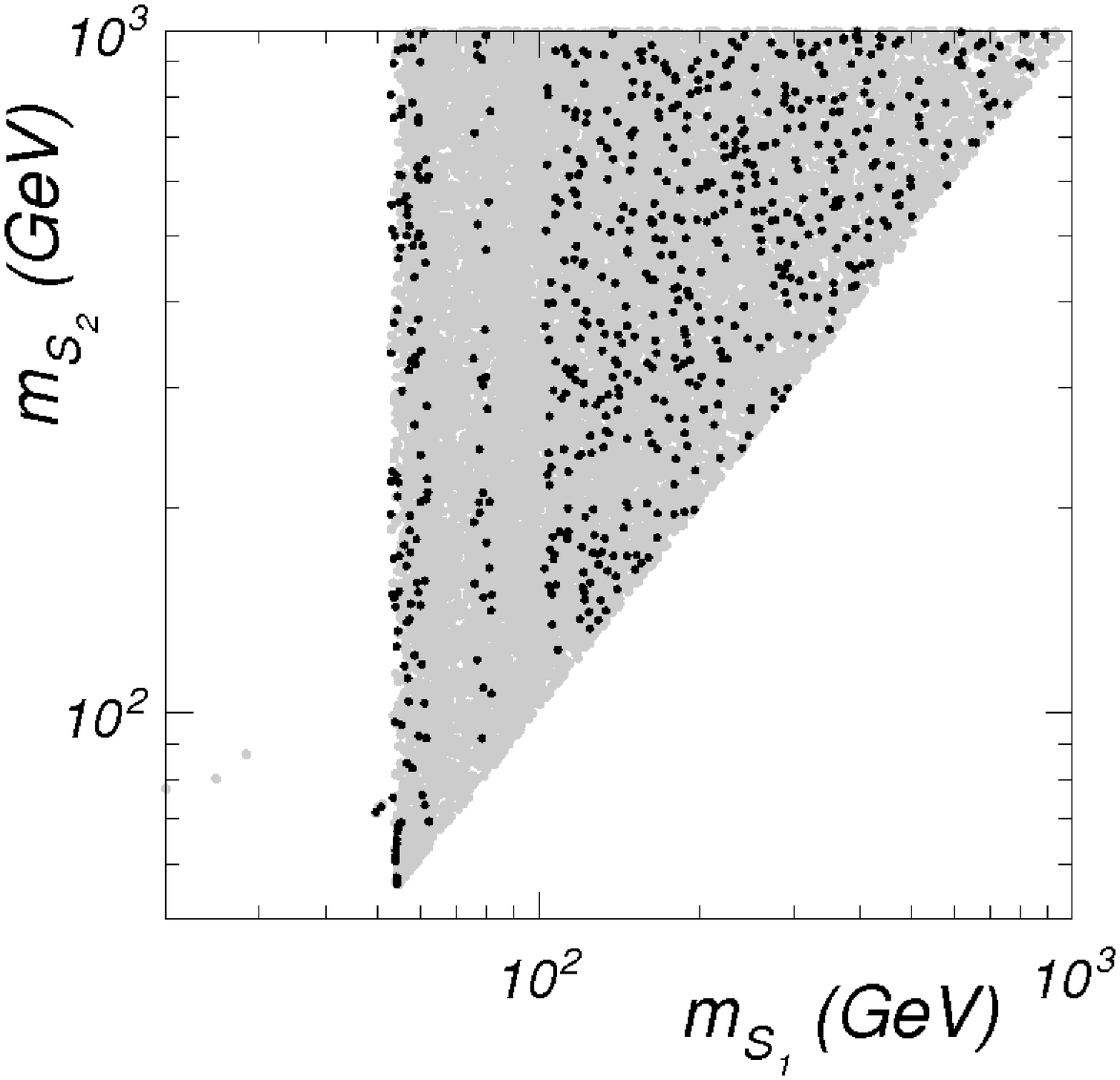} & \includegraphics[width=0.32\linewidth]{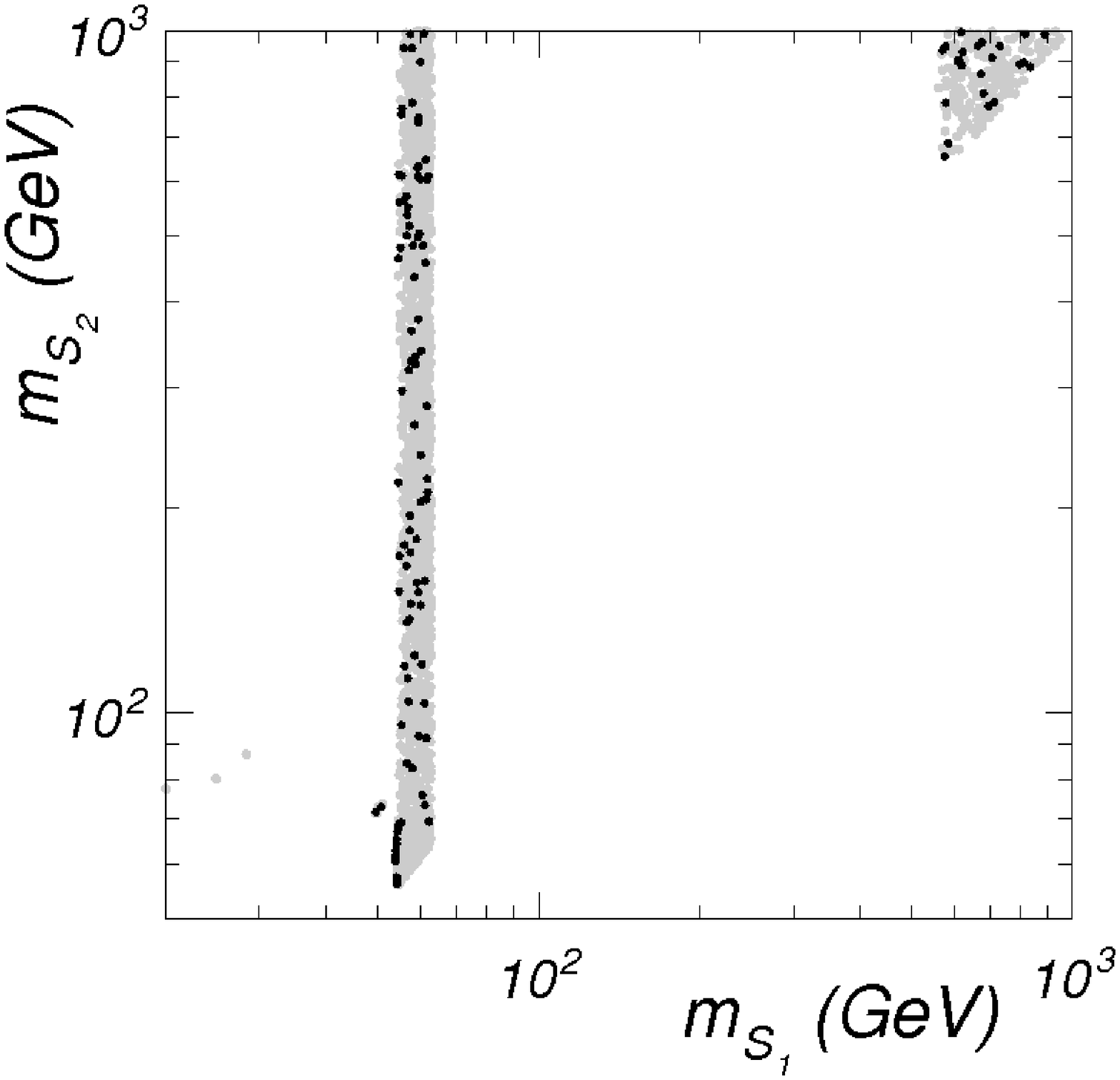} \\
\begin{sideways}$\qquad \qquad \qquad \lambda_{12}=$0.1  \end{sideways}& \includegraphics[width=0.32\linewidth]{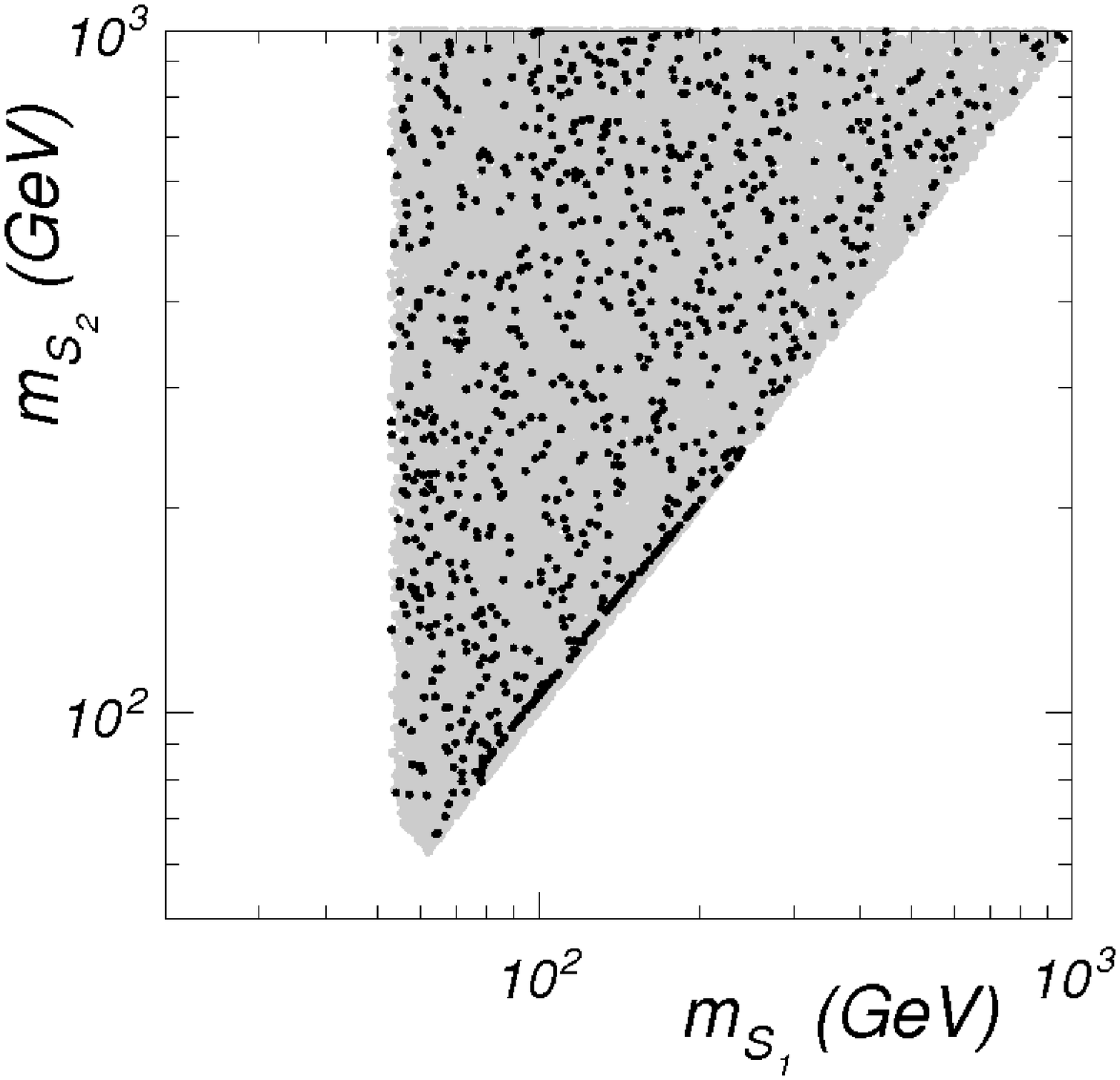} & \includegraphics[width=0.32\linewidth]{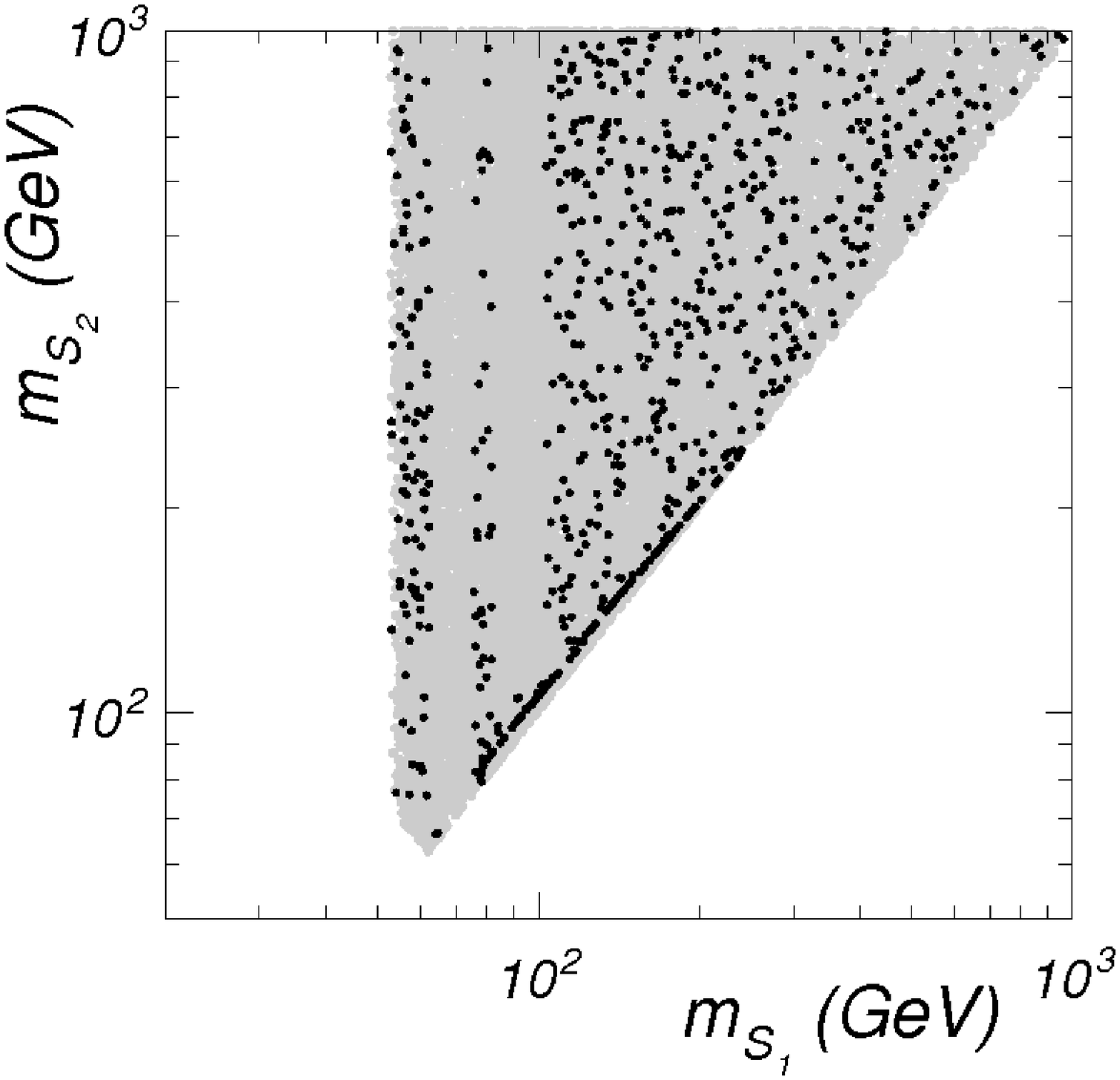} & \includegraphics[width=0.32\linewidth]{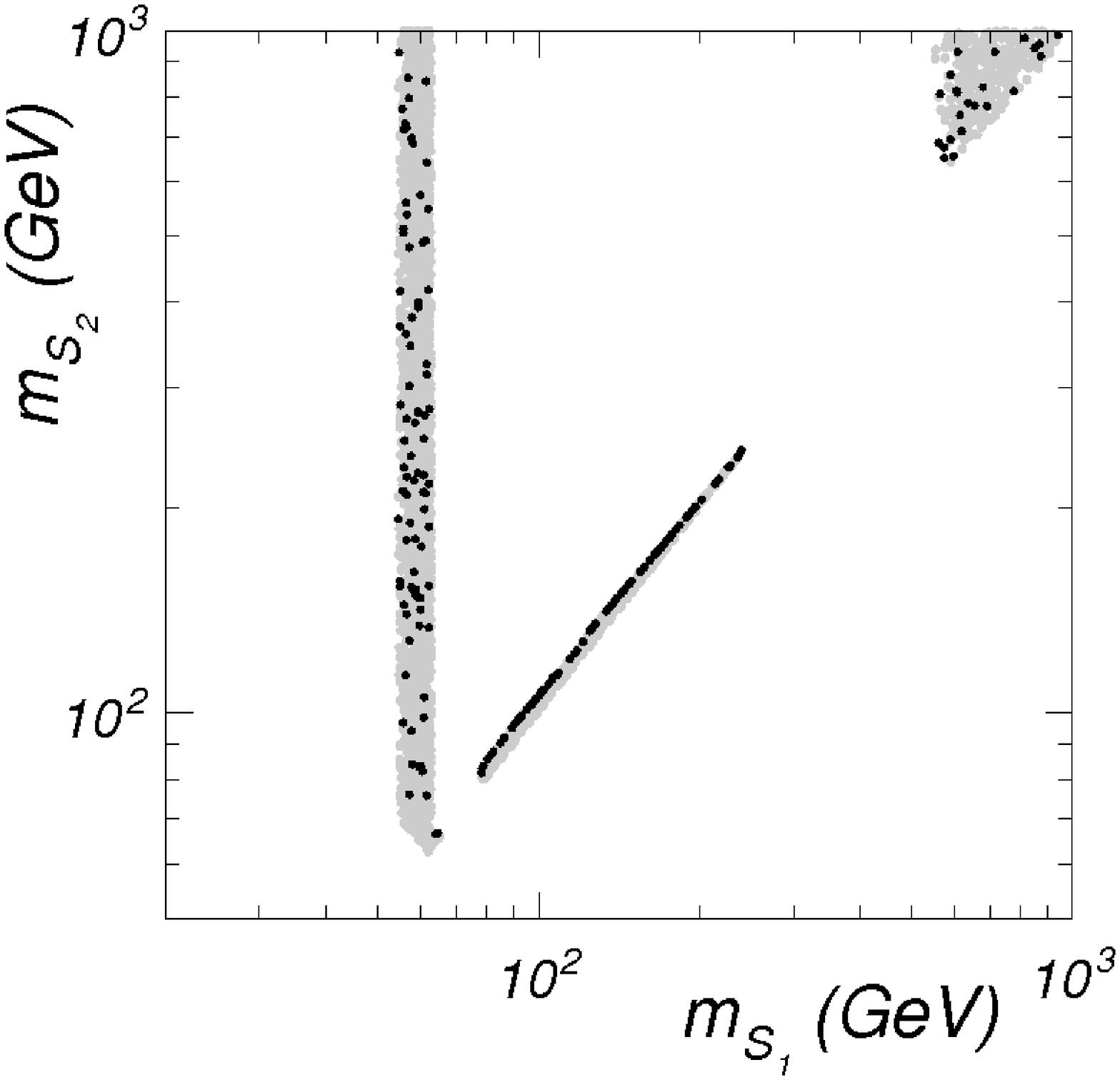} \\ 
\begin{sideways} $\qquad \qquad \qquad \lambda_{12}=1$  \end{sideways}& \includegraphics[width=0.32\linewidth]{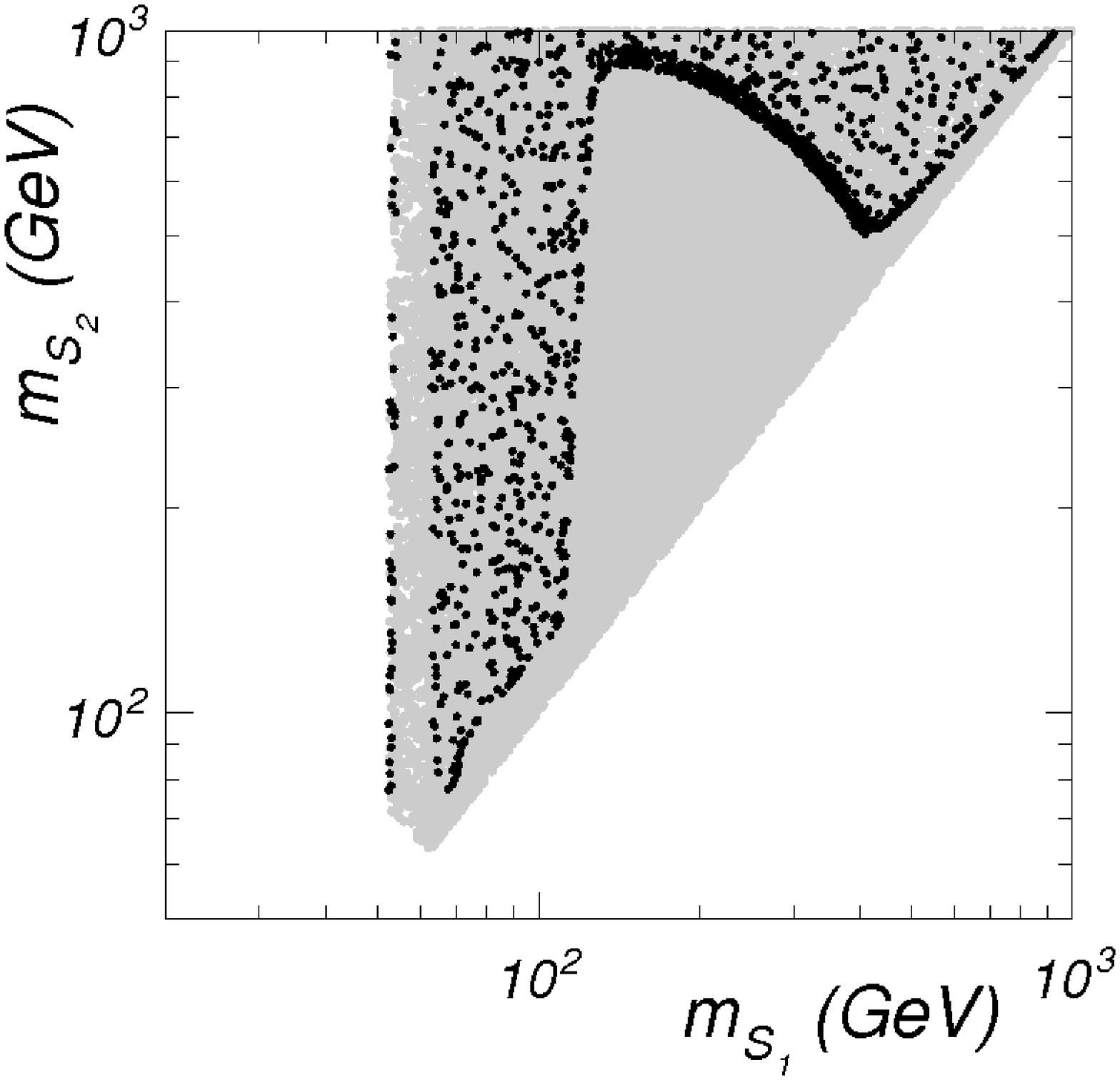} & \includegraphics[width=0.32\linewidth]{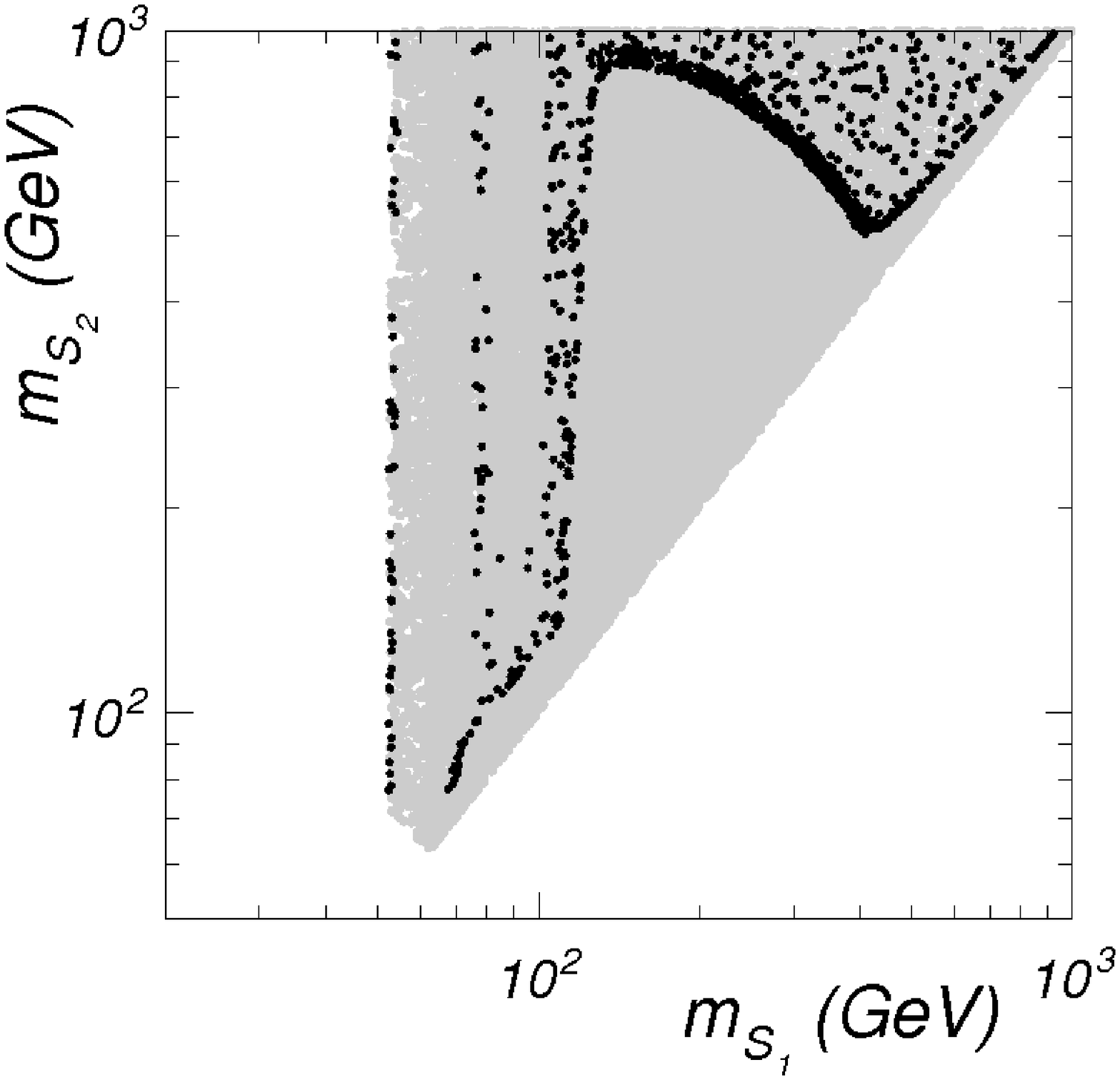} & \includegraphics[width=0.32\linewidth]{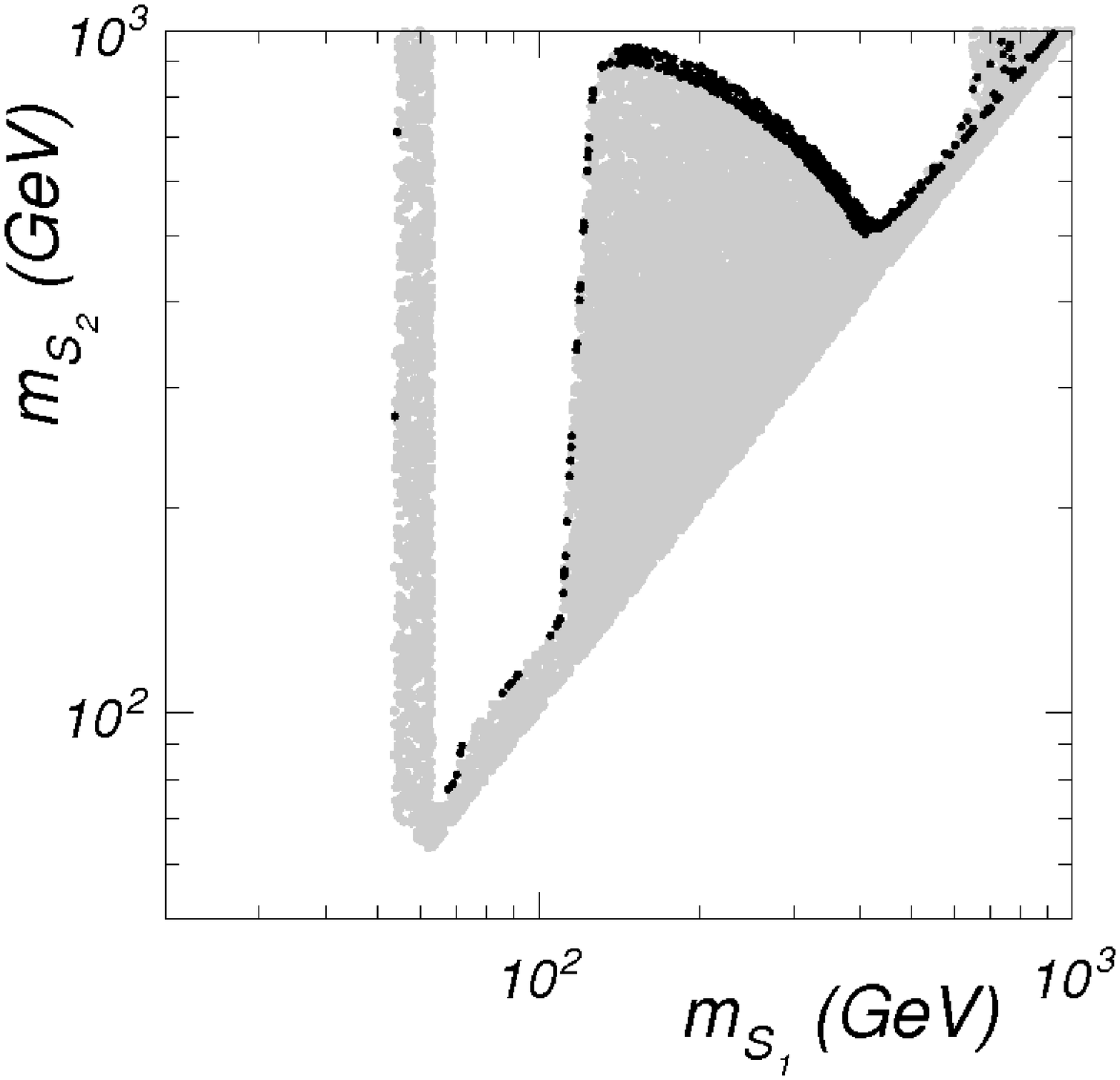} \\ 
\end{tabular}
\caption{
Effect of the experimental constraints in the $\{\msl,\,\msh\}$ parameter space of the ESHP model. We have used the same examples and colour conventions as in Fig.\,\ref{fig:l1m1}.}
\label{fig:m1m2}
\end{figure}

We have represented the results of the scans in Figs.\,\ref{fig:l1m1} and \ref{fig:m1m2}, where $\{\msl,\,\lambda_1\}$ and $\{\msl,\,\msh\}$ are plotted for fixed values of $\lambda_{12}$. From top to bottom, we have chosen $\lambda_{12}=0.01,\,0.1$, and $1$, respectively, thereby gradually switching on the effect of the extra singlet in the model. The different experimental constraints are added sequentially from left to right. The left column includes the bounds  from the invisible Higgs decay width and lifetime of $S_2$. The central column incorporates indirect detection bounds from Fermi-LAT results on the Galactic Centre and dSphs. Finally, in the right column we add the direct detection limits from LUX.
In all the plots, black dots correspond to those in which the (thermal) relic abundance of $S_1$ matches the results from the Planck satellite, whereas grey points are those in which $S_1$ is a subdominant dark matter component.

In all the plots of Fig.\,\ref{fig:l1m1} an accumulation of black dots along a thick line is visible, which coincides with the relic-density line of the standard SHP (the black line of Fig.\,\ref{fig:HPc}). For these points, the presence of the extra particle, $S_2$, has no effect, because the $\lambda_{12}$ coupling is too small or/and  $S_2$ is substantially heavier than $S_1$. These points appear as uniformly scattered in the $\{\msl,\,\msh\}$ plane in Fig.~\ref{fig:m1m2}. Besides this (somehow trivial) thick line, there are new regions of interest, which we discuss below.

The results for the top row ($\lambda_{12}=0.01$) resemble those of the usual SHP due to the smallness of $\lambda_{12}$. This can also be checked from the fact that the black dots in the plots in the first row of Fig.~\ref{fig:m1m2}, appear uniformly scattered in the allowed regions.
Consequently, the parameter space is extremely constrained by the combined effect of of the invisible Higgs width, indirect detection and (most notably) direct detection limits. Once all the bounds are included, only the points in the Higgs resonance and those with $\msl>500$~GeV survive. Still, when these results are compared to the left panel of Figure\,\ref{fig:HPc}, we observe a new (small) population of points at the Higgs resonance, with very small values of the coupling $\lambda_1$. This occurs when the masses of $S_2$ and $S_1$ are close enough so that coannihilation effects become important (first diagram of Fig.~\ref{fig:EHPprocesses}).  Away from the resonance region, the coannihilation effect is irrelevant due to the small size of $\lambda_{12}$ assumed here, so the correct relic density is obtained only for the usual value of $\lambda_1$, independently of how close $\msl$ and $\msh$ are.

As we increase the value of $\lambda_{12}$, new areas of the parameter space become available. In the middle row  of Fig.\,\ref{fig:l1m1}, ($\lambda_{12}=0.1$), we observe a region of black dots with masses $\msl\approx100-200$~GeV and a very small $\lambda_1$ coupling. These points have the correct relic abundance thanks to coannihilation effects, which requires $\msl\sim\msh$. They can be observed in the second row of Fig.\,\ref{fig:m1m2} as a thick line of black dots in that range of masses.

When $\lambda_{12}=1$ (last row of Fig.\,\ref{fig:l1m1}), the effect of the DM annihilation in two Higgses, $S_1 S_1\rightarrow hh$, exchanging $S_2$ in $t-$channel as in the last diagram of Fig.~\ref{fig:EHPprocesses}, becomes more remarkable, as soon as it is kinematically allowed, i.e. for $\msl\geq m_h$. This is the reason for the denser clouds of black dots out from the standard Higgs-portal thick line. For smaller values of $\msl$ co-annihilation is still the main responsible for DM annihilation, thus requiring the $S_1, S_2$ masses to be closer. All this can be seen in Fig.~\ref{fig:m1m2}. 
In the bottom panels of that figure we see that, for $\msl\leq m_h$, there is a thin ``black line" made of points close to $\msl=\msh$. The short distance of this line to the perfect degeneracy shows the required closeness between $\msl$ and $\msh$ to produce the amount of co-annihilation that gives the observed relic density. Below that line co-annihilation is too strong, so there are only gray dots (too low relic density).
For $\msl\geq m_h$ the line moves far away from $\msl=\msh$. As mentioned above, this behavior is due to the opening of the $S_1 S_1 \rightarrow h h$ process with both Higgses on-shell, which occurs via exchange of $S_2$ in $t-$channel (see Fig.~\ref{fig:EHPprocesses}). This process is very efficient, thus $\msh$ has to get much larger to appropriately decrease its effect and keep the relic density at the right value. However, as $\msl$ continues to increase, the black line again approaches $\msh\simeq \msl$. 
The reason is that the larger $\msl$ the less efficient the annihilation process, an effect that must be compensated in the $t-$channel diagram by a larger $\lambda_{12}$ or a smaller $\msh$; and the latter is the only possibility since we have set $\lambda_{12}=1$ in the plot. This can be easily understood by considering the $t-$channel diagram as generating an effective vertex, $S_1^2h^2$, with strength $\lambda_{\rm eff}\propto \lambda_{12}^2/\msh^2$.
In the next section we will elaborate more on this aspect.

\begin{figure}[t!]
\centering
\setlength{\tabcolsep}{0pt} 
\renewcommand{\arraystretch}{0} 
\begin{tabular}{c c c}
 $\lambda_{12}=$0.01  & $\lambda_{12}=$0.1  & $\lambda_{12}=1 $ \\ 
 \includegraphics[width=0.32\linewidth]{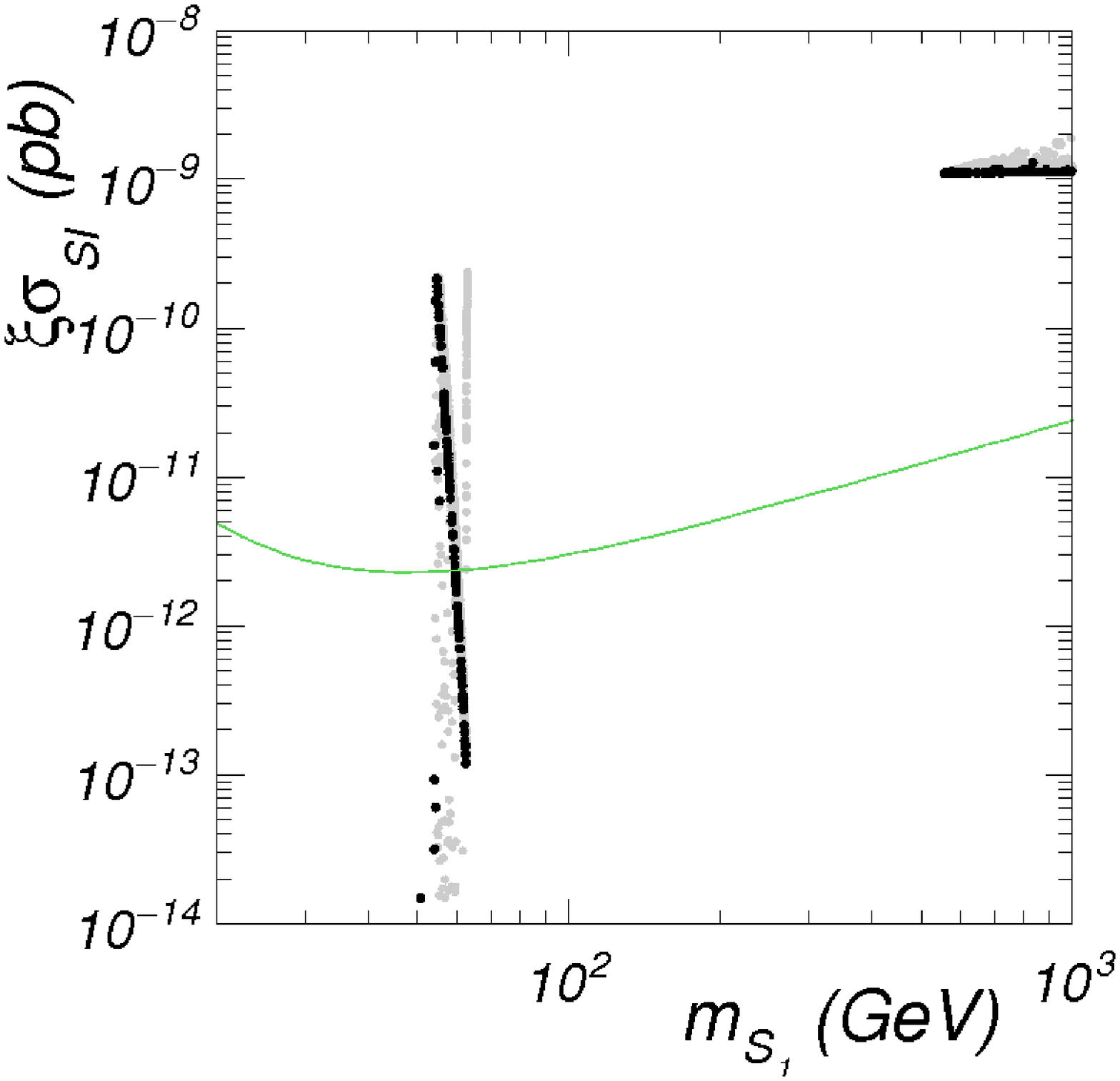} & \includegraphics[width=0.32\linewidth]{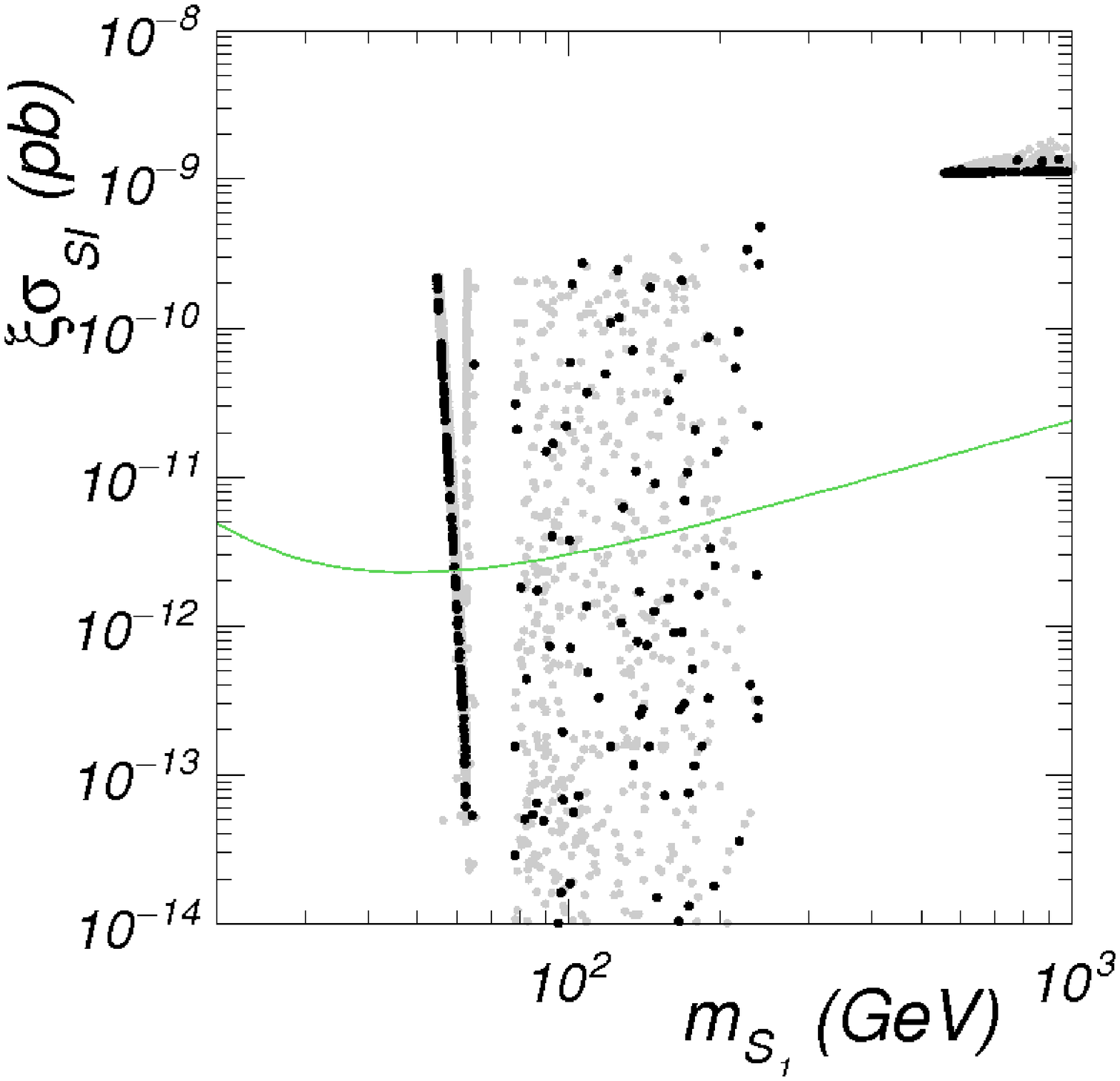} & \includegraphics[width=0.32\linewidth]{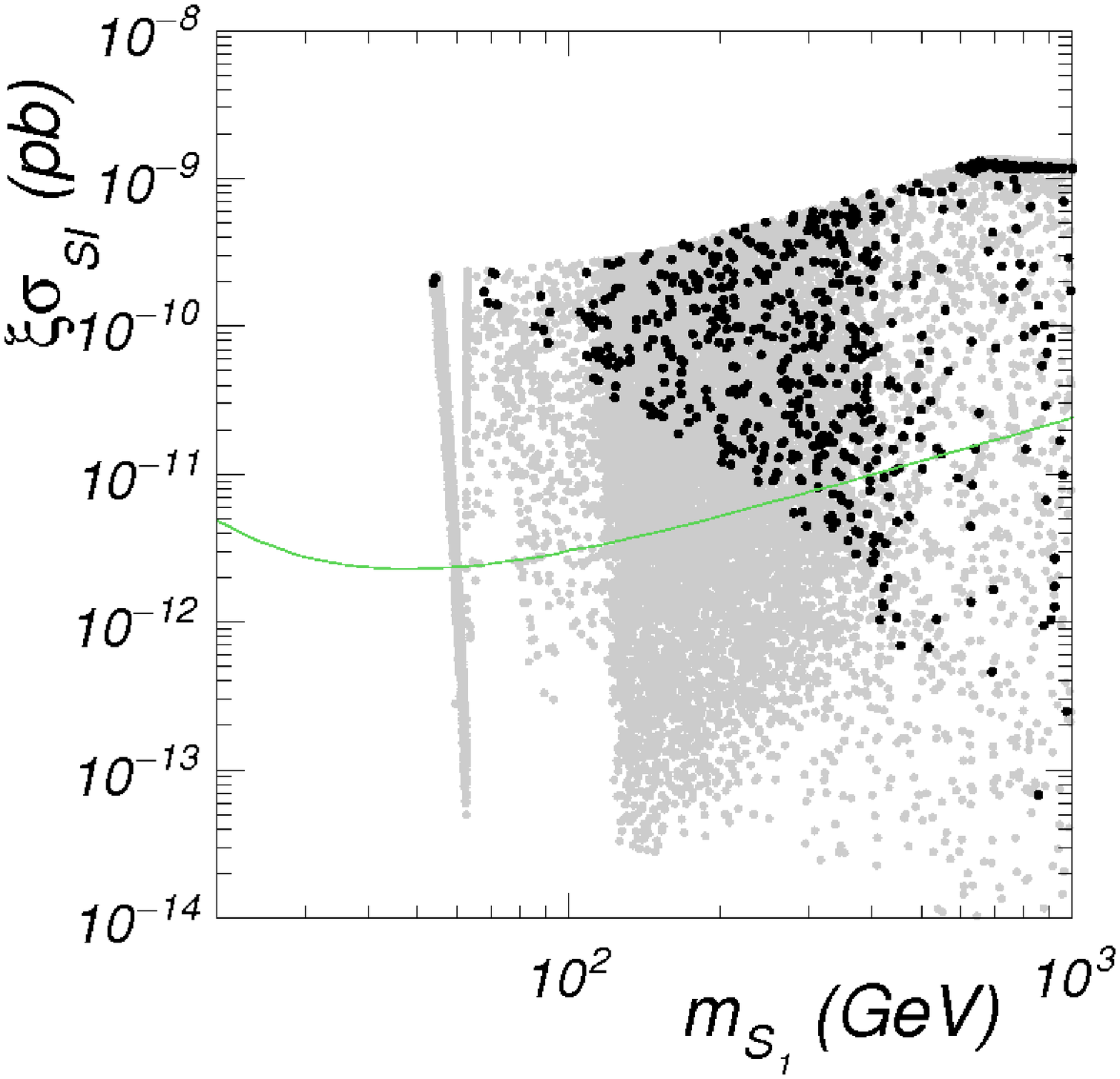} \\ 
\end{tabular}
\caption{Spin-independent scattering cross section of $S_1$ with protons as a function of its mass in the ESHP model. 
From left to right, we have fixed $\lambda_{12}=0.01,\,0,1$, and 1, respectively.}
\label{fig:lz}
\end{figure}

As in the case of the conventional SHP model, we expect future direct detection experiments (and in particular LZ) to be able to test large areas of the parameter space of our extended, ESHP, scenario. We represent in Fig.\,\ref{fig:lz} the theoretical predictions for the elastic scattering cross section of $S_1$ with protons, after all experimental constraints are applied. We indicate by means of a green line the expected reach of LZ. As we can observe, although a large area of the parameter space might be probed by these searches, there is a substantial region for which the predictions are beyond LZ sensitivity. For $\lambda_{12}=0.1-1$, this is possible for a range of DM masses between 100~GeV and 1~TeV (besides the usual narrow region at the Higgs resonance for $\msl\simeq m_h/2$), while satisfying the constraint on the relic abundance. None of these points can be probed by indirect detection either.

\section{Effective-theory description}
\label{sec:EFT}

As we have seen in the previous sections, the presence of the second particle, $S_2$, in the dark sector can enable the efficient annihilation of the DM particle, $S_1$, even if the usual quartic coupling of the latter, $\lambda_1 S_1^2 |H|^2$, is small enough to evade direct and indirect detection constraints. 

Since $\msh>\msl$, one can wonder whether $S_2$ might be integrated-out. Then, one would be left with a usual Higgs-portal scenario with just one particle, $S_1$, plus some higher-order operators, involving $S_1$ and $H$. If this procedure is sound, these additional operators should be ``clever" enough to mimic the effects of the heavy particle, $S_2$. Actually, the possibility of opening the allowed parameter-space of the Higgs-portal by adding new operators in the spirit of an effective field theory (EFT) has been considered in refs.~\cite{Brivio:2015kia, Fonseca:2015gva}. In our case, the coefficients of the EFT expansion are not completely independent, since they are determined by the ultraviolet (UV) completion, i.e., the Lagrangian of eq.(\ref{HPextlagr}). As we are about to see, this produces a quite special EFT, which is indeed very efficient in rescuing the excluded regions of the usual Higgs-portal for singlet scalar DM. Without the knowledge of the UV completion, such EFT could be seen as designed ad hoc for that purpose.

In fact, it is not always possible to mimic the effects of $S_2$ by integrating it out in some approximation. In particular, when $\msh\simeq\msl$, such integration is not appropriate. 
Consequently, the EFT description is not suitable to describe the regions of the parameter space where co-annihilation effects are dominant, e.g., for $\lambda_{12}\simlt 0.1$, see Figs.\,\ref{fig:l1m1} and \ref{fig:m1m2}.
However, there are other regimes in which $S_2$ is substantially (though not enormously) heavier than $S_1$, see for example Fig.\,\ref{fig:3examples} and the bottom row of Fig.~\ref{fig:m1m2}. In those cases the EFT captures, at least qualitatively, the relevant physics.

Once $S_2$ is integrated out at tree-level from eq.(\ref{HPextlagr}), the relevant new terms in the effective Lagrangian are
\begin{eqnarray}
\Delta\mathcal{L}_{\rm eff}(S_1,H)=-\frac{1}{2}
\frac{\lambda_{12}^2}{\msh^2}\ S_1^2 \left(|H|^2 - \frac{v^2}{2}\right)^2\ +\ \cdots .
\label{Leff}
\end{eqnarray}
Of course, this operator arises from the third tree-level diagram in Fig.\,\ref{fig:EHPprocesses}, with $S_2$ exchanged in $t-$channel.
Here the dots stand for higher order terms in $S_1$ or $H$. An important property of $\Delta\mathcal{L}_{\rm eff}$ is that, after EW breaking, the operator (\ref{Leff}) has the form $\frac{1}{4}S_1^2(h^2+2vh)^2$, triggering a contribution to the $S_1^2 h^2$ quartic coupling, without
generating new cubic couplings, $S_1^2 h$ (as a usual quartic coupling does). This is extremely useful to enhance the $S_1$ annihilation without contributing to direct-detection processes or to the Higgs invisible-width (if $S_1$ is light enough).

\begin{figure}[t!]
\centering 
\includegraphics[width=0.7\linewidth]{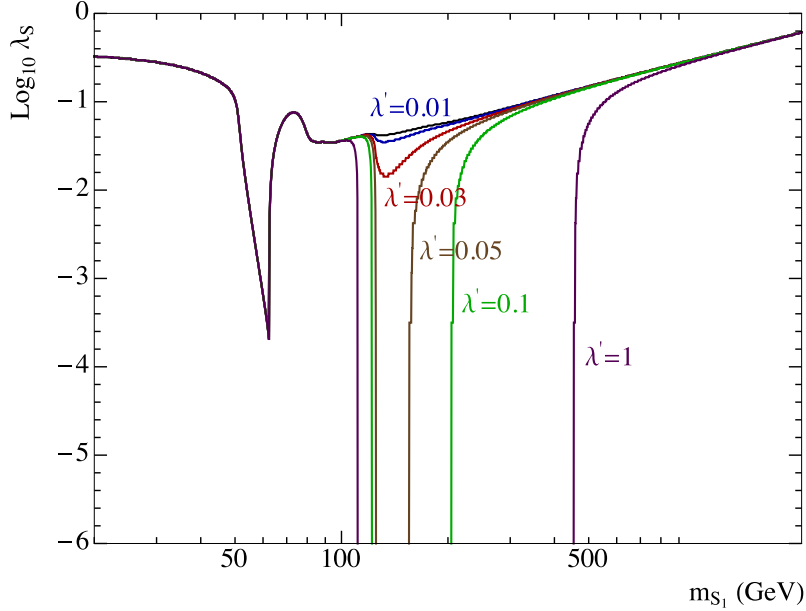}
\caption{Contour lines of the correct relic DM abundance
in an SHP effective theory consisting of the usual SHP Lagrangian plus an extra operator, as given in Eq.~(\ref{Leff2}), for several values of the $\lambda'$ coupling. This effective theory describes the ESHP in large regions of the parameter space. 
}
\label{fig:EFT}
\end{figure}

Fig.~\ref{fig:EFT} shows the performance of this Higgs-portal scenario with the presence of such extra operator, which we have parametrized as
\begin{eqnarray}
\mathcal{L}^{\ '}_{\rm SHP}=\mathcal{L}_{\rm SHP}
-\frac{1}{2}
\frac{\lambda'}{\msl^2}\ S_1^2 \left(|H|^2 - \frac{v^2}{2}\right)^2\ ,
\label{Leff2}
\end{eqnarray}
where ${\mathcal L}_{\rm SHP}$ is the SHP Lagrangian, defined in equation (\ref{HPlagr}), and $\lambda'=\lambda_{12}^2(\msl/\msh)^2$. The lines shown in the $\{\lambda_S,\,\msl\}$ plane correspond to the  correct relic abundance for different values of the effective coupling $\lambda^\prime$. As we can observe, the contribution from the effective operator triggers on when the annihilation channel into a pair of Higgs bosons gets kinematically allowed.  
Then, for a given value of $\msl$, 
as $\lambda^\prime$ increases, the value of $\lambda_S$ needed to recover the correct relic abundance decreases, eventually becoming irrelevant. For larger values of the DM mass, the effective operator becomes less efficient and eventually we recover the original behaviour. If we demand that $\lambda^{\prime}<1$, then the contribution from the effective operator is important for DM masses between 126~GeV and approximately 500~GeV. In this range of masses,
the usual quartic coupling $\lambda_S$ can be very small, thus helping to evade direct-detection limits. 

In other words, in this region of DM masses, for any value of the $\lambda_S$ coupling, there exists a value of $\lambda'$ that allows to recover the correct relic density. Since $\lambda'=\lambda_{12}^2(\msl/\msh)^2$, there are many combinations of the two underlying parameters of the U.V. theory, $\{\lambda_{12},\msh\}$, leading to the correct result.
These findings are in good agreement with the results presented in the previous section (Fig.\,\ref{fig:l1m1}), in particular with those for large $\lambda_{12}$ in the region of $\msl$, where the co-annihilation effects are not dominant.

\section{Conclusions}
\label{sec:conclusions}

One of the most economical and explored models of dark matter (DM) is the so-called singlet-scalar Higgs portal (SHP) model. It simply consists of an extra singlet scalar field (the DM particle), which is minimally coupled to the SM through interactions with the ordinary Higgs at the renomalizable level. Unfortunately, the experimental advances in direct and indirect dark matter searches, together with the latest results from the LHC, have ruled out vast areas of the viable parameter space of this scenario. Moreover, it is expected that future experiments will completely probe it within the next years and rule it out if no signal is found.

Motivated by the appealing simplicity of this model, we have considered in this article a minimal extension (ESHP) that could evade detection. It consists  of the addition of an extra real singlet scalar field in the dark sector, coupled also in a minimal, renormalizable way. 

We show that the new annihilation and/or co-annihilation channels involving the extra singlet allow to reproduce the correct relic abundance, even if the usual interaction of the DM particle with the Higgs were arbitrarily small. This allows to easily avoid the bounds from direct and indirect DM searches.

Apart from the DM mass and its coupling to the Higgs,
in its simplest version, the ESHP model has just two extra (relevant) parameters: the mass of the extra scalar and the quartic coupling between it, the DM particle and the Higgs field. Actually, the usual DM-Higgs coupling becomes irrelevant in most cases, since it is unnecessary, so the model has very few parameters. This permits to explore its phenomenology in an efficient way. In fact, though much more viable than the usual SHP model, this extended scenario is subject to a number of phenomenological constraints, most of them stemming from the mentioned quartic coupling between the DM, the extra scalar and the Higgs. These include bounds from the invisible width of the SM Higgs boson, the lifetime of the extra scalar particle, and direct and indirect searches for DM. Still, large portions of the parameter space survive all (present and even future) constraints.

We have also shown that, in the regions where the main extra effect is the annihilation of DM particles into SM particles (essentially Higgses), through the interchange in $t-$channel of the extra particle, the latter can be integrated-out, leaving a ``clever" SHP effective theory (just involving the DM particle and the Higgs) which can reproduce the relic density, while avoiding the usual strong constraints from DM searches. This is not possible however in the regions where the main extra effect is co-annihilation between the DM and the extra particle.

\appendix
\section{Radiative contributions to the $S_1 S_1 h$ vertex}
\setcounter{equation}{0}

In this appendix we compute the dominant radiative contributions for relevant physical processes involving DM in the context of the ESHP model, defined by the Lagrangian of eq.(\ref{HPextlagr}). We will do it in the framework of the EW-broken theory.

Assuming for simplicity and convenience a small $\lambda_2$ coupling, as has been done throughout the paper, the most important radiative corrections  are those contributing to the $S_1S_1 h$ vertex, in particular the three 1-loop diagrams depicted in Fig.~\ref{fig:1loopvertices}. This vertex plays a crucial for a number of DM processes; namely DM annihilation in the early universe, direct and indirect DM detection, and contributions to the invisible width of the Higgs boson.
Other relevant DM processes, in particular $S_1 S_1 \rightarrow hh$, receive radiative corrections as well, but they are much smaller than the contribution from the tree-level diagram in which a $S_2$ particle is exchanged in $t-$channel, see Fig.~\ref{fig:EHPprocesses}.

Therefore, in order to evaluate radiative corrections, the relevant terms of the Lagrangian in the broken phase are
\be
{\cal L } \supset   -\frac{1}{4!}\lambda h^4 - \frac{1}{3!} \lambda_1 v h^3 -\frac{1}{2}\lambda_{12}S_1S_2h^2 -  \lambda_{12} v S_1 S_2 h- \frac{1}{3!}\lambda_{31}S_1^3S_2\ .
\ee
In the following we will compute them, using the conventions of Ref.\,\cite{Peskin:1995ev} for Feynman rules.

Let us start with the one-loop diagrams involving two propagators (second and third diagrams of Fig.~\ref{fig:1loopvertices}). Their contribution to the vertex is given by
      \bea      
        \frac{iv}{16 \pi^2 }\left[  \lambda_{31}  \lambda_{12} B_0(p_h ^2; \msl, \msh)    +    \lambda_{12}^2 \left(B_0(p_{S_1} ^2; \msh, m_h) +B_0({p} ^2_{S_1'}; \msh, m_h)\right)\right]\ ,
       \eea
where $p_{S_1}$ and $p_{S_1'}$ represent the momenta of the two  $S_1$ particles entering the vertex, and
\be
B_0( p^2, m_1, m_2) = (\text{Divergent part}) + {\cal B}(p^2,  m_1, m_2)\ ,
\ee
with
\be
{\cal B}(p^2, m_1, m_2) = 
- \int^1_0 dx \log\frac{x m_1^2 +(1-x)m_2^2  - x(1-x) p^2}{m_1 m_2}\  .
\ee
In our case, the divergent part and the momentum-independent piece of ${\cal B}(p^2,  m_1, m_2)$ can be absorbed in the renormalized value of $\lambda_1$. Moreover, ${\cal B}(p^2,  m_1, m_2)$ can be expanded in powers of the momentum, as
\be
{\cal B}(p^2, m_1, m_2) =  1 - \frac{m_1^2 + m_2^2}{m_1^2 -m_2^2} \log\frac{m_1}{m_2} 
 +  p^2 F(m_1, m_2)  + {\cal O}(p^4) \ ,
 \label{B_expand}
\ee
with
\be
F(m_1, m_2) = \displaystyle  \frac{m_1^4 - m_2^4  - 2 m_1^2 m_2^2 \log\frac{m_1^2 }{m_2^2} }{2 (m_1^2 - m_2^2)^3}\ .
\ee

Keeping just the term proportional to $p^2$  turns out to be a good approximation in most cases (recall here that the $p-$independent terms in Eq.~(\ref{B_expand}) are absorbed in a finite renormalization of $\lambda_1$).  Hence a good approximation for the contribution to the $S_1S_1h$ vertex from the one-loop diagrams involving two propagators is
      \bea
           \Gamma^{(2)} \simeq       \frac{iv}{16 \pi^2 }\left[  \lambda_{31}  \lambda_{12} p_h ^2 F(\msl, \msh)    +    \lambda_{12}^2( p_{S_1} ^2 + {p} ^2_{S_1'})F(\msh, m_h) \right]\ .
      \eea
        
Alternatively, this contribution to the vertex van be viewed as the Feynman rule stemming from the corresponding term in the effective action, namely
\be
16 \pi^2\Delta^{(2)}{\cal L} =     
-  \frac{1}{2} \lambda_{31}  \lambda_{12} v  F(\msl, \msh)   \; S^2  \partial^2  h
 -  \lambda_{12}^2  v \,  F(\msh, m_h)   \;   S ( \partial^2  S)    h \ .
\ee
This is a convenient way to encode these contributions in the MicrOMEGAs code, as we have done throughout the paper.

Let us now consider the one-loop diagrams involving three propagators (fourth diagram of Fig.~\ref{fig:1loopvertices}). The main difference with the previous two diagrams is that this represents a finite contribution which should be entirely counted, even the momentum-independent contribution, since the latter corresponds to a $S_1^2 |H|^4$ operator in the unbroken theory and cannot be absorbed in a finite renormalization of $\lambda_1$. Using the same momentum expansion as before, the corresponding contribution to the $S_1S_1h$ vertex reads
      \bea
           \Gamma^{(3)}  &\simeq&  \frac{i}{16 \pi^2 }\lambda_{12}^2  \lambda   v^3    \left[ F_3(\msh, m_h,m_h)  
             \right.\\  &&    \left.
                    +  (p_{S_1}^2+p_{S_1'}^2) G(\msh, m_h,  m_h ) 
                    +   p_h^2 G(m_h , m_h , \msh ) \right],\;\; \nonumber \\   
       \eea
with
\bea\begin{aligned}
F_3 (m_1, m_1,m_2)  =   \;   
          &- \frac{  m_1^2 - m_2^2  -m_2^2  \log{\frac{m_1^2}{ m_2^2}}}{(m_1^2 -m_2^2)^2}\ ,  \\[3mm]
G(m_1,m_1,m_2)=  \;   
         & - \frac {m_1^6 -6 m_1^4 m_2^2  + 3 m_1^2 m_2^4  +2 m_2^6 +  6 m_1^2  m_2^4  \log \left( \frac{m_1^2}{m_2^2} \right) }
                   {12 m_1^2 (m_1^2-m_2^2)^4}   \; \  ,
          \\[2mm]     
G(m_1,m_2,m_1) =  \;   
         & - \frac{m_1^4 + 4 m_1^2 m_2^2 - 5 m_2^4 - 2 m_2^2 (2 m_1^2 + m_2^2) \log \left( \frac{m_1^2}{m_2^2} \right) }
                   {4 (m_1^2-m_2^2)^4}\ .
\end{aligned}\eea
The corresponding terms in the effective action read
\bea
16 \pi^2\Delta^{(3)}{\cal L} &=&    \frac{1}{2}  \lambda_{12}^2  \lambda v^3
              \left[  \phantom{  \frac{a}{b}}   \!  \!  \!  \!  \!  \right.   
                          \; F_3(\msh, m_h,  m_h)  \; S^2 h  \nonumber \\[3mm]
  &   &        \left.          - 2 G(\msh, m_h,  m_h ) S ( \partial^2  S)    h  
                 -   G(m_h , m_h , \msh )   \;   S^2  \partial^2 h \phantom{  \frac{a}{b}}   \right]         \ .
\eea 

\vspace{0.2cm}
\noindent{\bf \large Acknowledgements}

This work has been partially supported by the MICINN / MINECO, Spain, under contracts FPA2013-44773-P, FPA2015-65929-P and FPA2016-78022-P. We also thank the Spanish MINECO Centro de excelencia Severo Ochoa Program under grant SEV-2012-0249, the Consolider-Ingenio CPAN CSD2007-00042, as well as MULTIDARK CSD2009-00064. 
D.G.C acknowledges support from the STFC and the partial support of the Centro de Excelencia Severo Ochoa Program through the IFT-UAM/CSIC Associate programme. The work of J.Q. is supported through the Spanish FPI grant SVP-2014-068899. J.Q. thanks Sandra Robles for her invaluable help with computational resources.

\providecommand{\href}[2]{#2}\begingroup\raggedright\endgroup

\end{document}